\documentclass[12pt]{article}

\usepackage{scicite}

\usepackage{times}

\usepackage{graphicx} 
\usepackage{xcolor} 

\newenvironment{sciabstract}{%
\begin{quote} \bf}
{\end{quote}}

\title{Gapped magnetic ground state in quantum-spin-liquid candidate
$\kappa$-(BEDT-TTF)$_2$Cu$_2$(CN)$_3$}

\author
{Bj{\"o}rn Miksch,$^{1}$ Andrej Pustogow,$^{1,2}$ Mojtaba {Javaheri Rahim},$^{1}$\\ Andrey A. Bardin,$^{3}$ Kazushi Kanoda,$^{4}$ John~A.~Schlueter,$^{5}$\\ Ralph H{\"u}bner,$^{1}$ Marc Scheffler,$^{1}$ Martin Dressel$^{1\ast}$\\
\\
\normalsize{$^{1}$1.~Physikalisches Institut, Universität Stuttgart,}\\
\normalsize{Pfaffenwaldring 57, 70569 Stuttgart, Germany}\\
\normalsize{$^{2}$Department of Physics and Astronomy, UCLA,}\\
\normalsize{Los Angeles, California 90095, USA}\\
\normalsize{$^{3}$Institute of Problems of Chemical Physics, Russian Academy of Sciences,}\\
\normalsize{Chernogolovka, Moscow Region 142432, Russia}\\
\normalsize{$^{4}$Department of Applied Physics, University of Tokyo, Tokyo 113-8656, Japan}\\
\normalsize{$^{5}$Material Science Division, Argonne National Laboratory, Argonne, Illinois 60439-4831 and}\\
\normalsize{National Science Foundation, Alexandria, Virginia 2223, USA}\\
\\
\normalsize{$^\ast$To whom correspondence should be addressed; E-mail:  dressel@pi1.physik.uni-stuttgart.de.}
}

\date{}

\begin{document}


\baselineskip24pt


\maketitle


\begin{sciabstract}
  Geometrical frustration, quantum entanglement and disorder may prevent
  long-range order of localized spins with strong exchange interactions, resulting
  in a novel state of matter. $\kappa$-(BEDT\--TTF)$_2$\-Cu$_2$(CN)$_3$ is considered the best
  approximation of this elusive quan\-tum-spin-liquid state, but its ground-state
  properties remain puzzling. Here we present a multi-frequency electron-spin resonance
  study down to millikelvin temperatures, revealing a rapid drop of the
  spin susceptibility at $T^*=6\,$K. This opening of a spin gap, accompanied by
  structural modifications, suggests the enigmatic `6~K-anomaly' as the transition
  to a valence-bond-solid ground state. We identify an impurity contribution
  that becomes dominant when the intrinsic spins form singlets. Only probing
  the electrons directly manifests the pivotal role of defects for the low-energy
  properties of quantum-spin systems without magnetic order.
\end{sciabstract}


The exotic properties of quantum spin liquids (QSL) have continuously drawn interest since Anderson’s seminal study half a century ago~\cite{Anderson73} where he considered spin models that possess an extensive degeneracy of states.
While classical spins in magnetically interacting systems mostly end up in a long-range periodic arrangement, it is widely believed that geometrical frustration may suppress conventional magnetic ordering down to $T=0$, giving rise to a unique, fluctuating, quantum-disordered state~\cite{Balents10,Savary16,Broholm2020}.
Organic charge-transfer salts were the first and most versatile QSL candidates because their microscopic parameters can be easily tuned by chemical means.
They crystallize in a near-isotropic triangular arrangement of $S=\frac{1}{2}$ spins on molecular dimers~\cite{Powell11} (Fig.~1), in contrast to most inorganic QSL candidates, such as pyrochlore compounds or Herbertsmithite, which form tetrahedral or kagome lattices~\cite{Gingras14,Norman16}, respectively.

For two decades QSLs have been intensely explored by various magnetic probes, but for most materials crucial questions remain unanswered:
How is magnetic order prevented?
What is the ground state?
And what is the spin excitation spectrum?
For the two-dimensional charge-transfer salts, the importance of disorder became evident, lately~\cite{Furukawa15b,Saito18,Riedl19,Kawamura19,Pustogow19}.
On the fundamental issue of a spin gap in $\kappa$-(BEDT\--TTF)$_2$\-Cu$_2$(CN)$_3$, however, conflicting conclusions are given by magnetic torque~\cite{Isono16}, muon spin rotation ($\mu$SR)~\cite{Pratt11}, thermal transport~\cite{Yamashita09}, specific heat~\cite{Yamashita08} and nuclear-magnetic-resonance (NMR) measurements~\cite{Shimizu03}.
The necessity to study the range $T\rightarrow 0$ favored experimental methods that are susceptible also to impurity spins.
While the bulk magnetization, measured by SQUID or torque magnetometry, does not distinguish between intrinsic or extrinsic contributions, $\mu$SR and NMR spectroscopies are indirect probes as they record the influence of the local magnetism on the spectral and relaxation properties of muons and atomic nuclei, respectively.
Electron spin resonance (ESR), on the other hand, directly probes the magnetic excitation spectrum of the conduction electrons, which allows us to unambiguously identify the intrinsic response and separate it from other contributions.
To that end, we developed a novel broadband ultra-low-temperature ESR technique that exceeds common limits of this method.

In Fig.~2A the temperature dependence of the spin susceptibility $\chi_S(T)$ is plotted, as derived from X-band ESR measurements on $\kappa$-(BEDT\--TTF)$_2$\-Cu$_2$(CN)$_3$ single crystals.
The overall behavior can be described by a Heisenberg model on a triangular lattice with strong antiferromagnetic exchange interaction $J=250\,$K, in agreement with previous estimates~\cite{Shimizu03}.
However, at $T^*\approx 6\,$K a rapid drop in $\chi_S(T)$ is observed, the very same temperature where an anomaly was consistently identified by various methods~\cite{Yamashita09,Yamashita08,Shimizu03,Manna10,Poirier14,Shimizu06}.
Fitting the decay by an activated behavior $\chi_S(T) \propto T^{-1}\exp\{-\Delta/T\}$ yields a spin gap of $\Delta = 12.1\,$K, as shown by the green line; details are given in~\cite{SM}.
Since the $g$-value of $\kappa$-(BEDT\--TTF)$_2$\-Cu$_2$(CN)$_3$ remains unaffected (Fig.~S2), long-range magnetic order or any well-defined local moments can be ruled out, in accord with previous NMR results that show no splitting in the spectra~\cite{Shimizu03}.
Thus, our ESR investigations unambiguously identify the anomaly at $T^*$ as a phase transition to a gapped magnetic ground state.

Because our findings clearly rule out the widely assumed gapless QSL state with itinerant spinons~\cite{Yamashita08,Shimizu03}, let us consider
possible alternative scenarios for a spin-gapped ground state on a slightly distorted triangular lattice (Fig.~1B), such as a valence bond solid (VBS), an Amperean  pairing instability,
$Z_2$ QSL  or other resonating valence bond phases \cite{Balents10,Savary16,Zhou17,Broholm2020,Kimchi18}.
At first, we notice that $\chi_S(T)$ resembles other known systems that undergo a continuous transition to a spin-gapped state, for instance the well-known spin-Peierls
transitions in organic linear-chain compounds~\cite{Dumm00a}, or inorganic CuGeO$_3$ and $\alpha^{\prime}$-NaV$_2$O$_5$~\cite{Hase93,Isobe96}, as elaborated in the Supplementary Materials~\cite{SM}.
Similar to these quasi 1D systems, also $\kappa$-(BEDT\--TTF)$_2$\-Cu$_2$(CN)$_3$ exhibits a structural anomaly with anisotropic thermal expansion at the transition~\cite{Manna10}, corroborating the idea of a broken-symmetry ground state that couples to the lattice.
Such a transition occurs when the energy gain by the formation of spin singlets exceeds the energy required for the lattice distortion.
Here, the shrinkage of the $c$-axis below $T^*$ that is accompanied by pronounced lattice softening~\cite{Poirier14} suggests that the $(b\pm c)$-directions are the preferable orientations (Fig.~1C).
Taken together, these experimental signatures are fully consistent with a VBS as the nonmagnet ground state -- a situation also discussed towards kagome~\cite{Norman2020} and higher-dimensional QSL candidates~\cite{Hermanns15}.

Besides the VBS scenario, a gapped QSL phase could be the result of a topological $Z_2$ spin liquid found in perfect triangular-lattice dimer models with some analogy to a phase-disordered BCS superconductor~\cite{Broholm2020,Savary16,Moessner01}.
However, the present in-plane anisotropy and potential symmetry breaking at $T^*$ put tight constraints on a conceivable $Z_2$ state.
Alternatively, it was suggested that an Amperean pairing instability can impose a gap to mobile spinons, with an incommensurate modulation of the amplitude~\cite{Lee07}, in order to explain the phase transition at $T^*$ and other low-temperature properties of the title compound.
Yet, the latter scenario is rather difficult to reconcile with the vanishing thermal transport~\cite{Yamashita09} and presence of unscreened orphan spins, discussed below.
Precise structural studies through $T^*$ may prove decisive to distinguish between the above scenarios by deducing the valence bond arrangement.

Having clarified the ground state as a nonmagnetic spin-singlet phase -- why did it remain concealed for decades despite all efforts?
To that end, let's take a careful look at the ESR raw data in Fig.~2B.
Below room temperature, the Dysonian absorption at $B_\textrm{main}\simeq 337\,$mT acquires a Lorentzian shape as the conductivity decreases in accordance with dc transport measurements~\cite{Pinteric14}.
The line initially broadens when cooled down to $40\,$K, followed by a moderate reduction of the line width $\Delta B$ at lower temperatures (Figs.~S2 and S3 in~\cite{SM}).
Near $T^*$ the signal narrows extremely and the doubly integrated area (corresponding to $\chi_S(T)$) is strongly reduced.
Most important, a second component with even smaller $\Delta B$ appears at $T^{\ast}$, as illustrated in Fig.~2C.
While slightly below $T^{\ast}$ the resonance field of the two features is indistinguishable, panel (D) shows how the newly emerging component splits off as an additional peak that shifts away from $B_\textrm{main}$ for $T<2.5\,$K.
We assign this contribution with increasing intensity upon cooling (orange symbols in Fig.~2A) to defects, in accord with previous considerations~\cite{Riedl19,Kawamura19,Pustogow19,Shimizu06}.
The feature actually consists of 2--3 individual lines with slightly different angle dependence and field variation, consistently observed in all crystals from four different laboratories;
while the type of defects is always the same, the density varies.
As they originate from the main ESR line at $B_\textrm{main}$, we conclude that these defect spins arise from molecular dimer sites that remain unpaired below $T^*$.

Fig.~2E illustrates the low-temperature angle dependence of the ESR lines within the $bc$-plane.
The signal at $B_\textrm{main}$ shifts only by $10^{-3}$ upon rotation, caused by the anisotropic $g$-tensor of the BEDT-TTF molecules due to spin-orbit interactions.
The minor lines, however, exhibit an order of magnitude larger variations and follow a $(3\cos ^2{\theta} -1)$ angular dependence, implying that the (BEDT-TTF)$_2^+$ defect spins are subject to dipole-dipole interaction with a nearby magnetic moment ($r\approx 6$--$7\,$\AA), as suggested recently~\cite{Pustogow19}.
We consistently observe similar angle dependences in different crystals indicating that the defect spins are localized and experience dipolar interaction with a local magnetic moment along a fixed crystallographic direction, in particular ruling out that this local field is caused by regular dimers within the $bc$-plane.
In all samples, however, we detect $S=1/2$ Cu$^{2+}$ ions in the anion sheet (Fig.~S6)~\cite{Komatsu96,Padmalekha15}, which can generate comparable local fields at the organic dimer sites~\cite{SM}.
These excess electrons most likely dope the closest (BEDT-TTF)$_2$ site creating a vacancy and, hence, an unpaired spin on one of the neighboring dimers (sketched in Fig.~1C).
Thus, Cu$^{2+}$ impurities may be responsible for the observed defect spins with $(3\cos ^2{\theta} -1)$ ESR signal.
Full clarification of the nature and origin of the local magnetic moments remains a desideratum
for experiment as well as theory.

The emergent local fields for $B\parallel a^*$ are of comparable strength as the low-temperature $^{13}$C NMR line width ($4.8\,$mT)~\cite{Shimizu06} and signatures in the $\mu$SR data~\cite{Pratt11}.
The distinct anisotropy also explains the angular shift and diverging susceptibility in the magnetic torque observed for low $T$ and $B$~\cite{Isono16,Riedl19}.
To elucidate the relation to the `weak-moment antiferromagnetic phase' suggested in Ref.~\cite{Pratt11}, we performed broadband ESR experiments at different fields down to mK-temperatures utilizing superconducting coplanar waveguide resonators as illustrated in Fig.~3A.
As an example, Fig.~3B displays the temperature evolution of the ESR absorption of the crystal measured with the fundamental mode at $1.1\,$GHz upon cooling from $4\,$K to $25\,$mK.
The defect signal, affected by local moments, separates from $B_\textrm{main}=40\,$mT at $T_{\rm loc}\simeq 1\,$K and saturates at lower fields upon cooling;
the field dependence of $T_{\rm loc}$ is in accord with the suggested phase boundaries~\cite{Pratt11}.
Fig.~3C illustrates the approximately field-independent offset of the defect signal with respect to $B_\textrm{main}$ at $T_\textrm{base}=25\,$mK.

What is the origin of the second line in Fig.~3B at $B_\textrm{main}$?
While close to $T^*$ thermal excitations across the spin gap exceed the defect contributions (Fig.~2), they should not contribute at $T_\textrm{base}\simeq \Delta/500$.
As sketched in  Fig.~1C, there is a possibility of intrinsic valence-bond imperfections, for instance through domain walls or other types of broken singlets~\cite{Riedl19,Kawamura19,Kimchi18}.
In absence of a nearby magnetic moment, the corresponding ESR line remains at $B_\textrm{main}$.

There is an obvious advantage of using electron spins to directly probe the magnetic properties of quantum spin liquids.
Since the NMR spin-lattice relaxation rate is susceptible to any kind of unpaired spins in the sample, it will be dominated by impurities in case a spin gap opens.
Indeed, a recent field-dependent NMR study concluded upon polarization of the defect spins for fields $B\simeq 10\,$T in several $\kappa$-type organic QSL compounds~\cite{Pustogow19}.
Of course, impurities like Cu$^{2+}$ do not dissolve when warming above $T^*$, but they are overwhelmed by the large number of intrinsic paramagnetic moments.
While high densities of Cu$^{2+}$ will dope the system into a metallic state~\cite{Komatsu96}, tiny amounts of charged defects imbedded in a Mott-insulating matrix are a potential source of electrical polarization, possibly accounting for the controversially discussed relaxor-like dielectric response~\cite{Abdel10,Pinteric14}.

The scenario of localized unpaired spins, possibly pinned to Cu$^{2+}$, dispersed in a VBS resolves the longstanding controversy of a vanishing $\kappa/T$ in thermal transport for $T\rightarrow 0$ despite the gapless excitations concluded from specific heat~\cite{Kawamura19,Yamashita09,Yamashita08}.
Recent reexamination of the thermal conductivity suggests a spin gap also in $\beta^{\prime}$-EtMe$_3$\-Sb[Pd(dmit)$_2$]$_2$~\cite{Bourgeois-Hope2019}.
While much larger disorder effects are expected for the inorganic QSL candidate Herbertsmithite ZnCu$_3$(OH)$_6$Cl$_2$ due to Cu-Zn antisite exchange of order $10\,$\%~\cite{Freedman2010}, which indeed have been reported~\cite{Fu15,Khuntia20}, Ag$^{2+}$ defects should be absent in $\kappa$-(BEDT\--TTF)$_2$\-Ag$_2$(CN)$_3$.
Nevertheless, defects spins prove crucial for the low-temperature magnetic properties of all quantum spin systems that lack magnetic order.
There are few QSL candidates remaining where the opening of a spin gap has not been proven beyond any doubt, but maybe this is just a question of time.
Our newly developed broadband low-$T$ ESR spectroscopy now provides a versatile tool to tackle
these and related issues.



\bibliography{kCuCN_EPR}

\bibliographystyle{Science}

\section*{Acknowledgments}
We thank S.E. Brown, K. Holczer, R.K. Kremer, G. Gorgen Lesseux and S.M. Winter for fruitful discussions.
{\bf Funding:} The work at Stuttgart was supported by the Deutsche Forschungsgemeinschaft via DR228/39-3. A.P. acknowledges support by the Alexander von Humboldt Foundation through the Feodor Lynen Fellowship. J.A.S acknowledges support from the Independent Research and Development program from the NSF while working at the Foundation and from the National High Magnetic Field Laboratory (NHMFL) User Collaboration Grants
Program (UCGP). {\bf Author contributions:} M.D. conceived the project; M.S. designed the low-temperature ESR facilities; B.M. and M.J.R. performed the experiments; B.M. and A.P. analyzed and interpreted the results, in perpetual exchange with M.D.; A.A.B, K.K., J.A.S. and R.H. are responsible for the crystal growth; B.M., A.P. and M.D. wrote the manuscript with contributions from other authors.
{\bf Competing interests:} The authors declare that they have no competing financial interests.

\section*{Supplementary materials}
Materials and Methods\\
Supplementary Text\\
Figs. S1 to S14\\
References \textit{(39 - 74)}


\clearpage
\begin{figure}
\includegraphics[width=\textwidth]{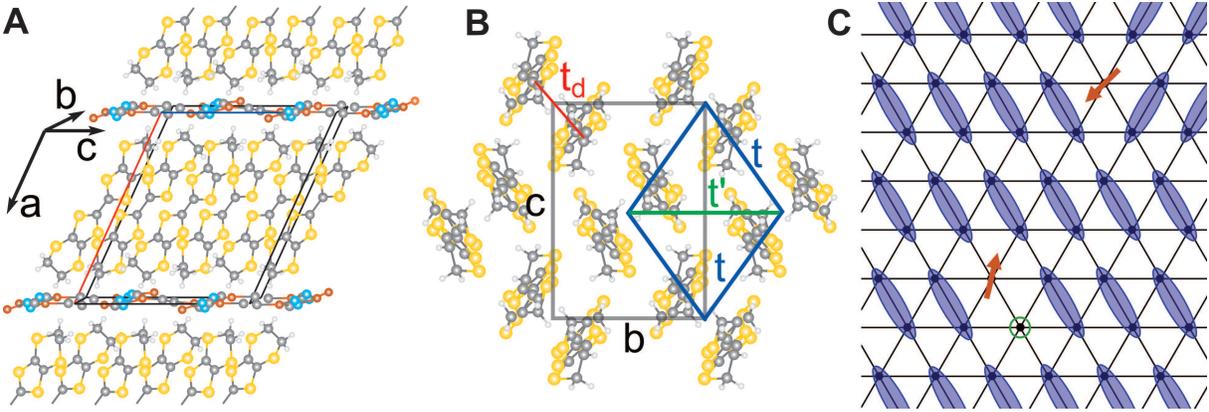}
\caption{
\textbf{Crystal structure of the QSL candidate $\kappa$-(BEDT\--TTF)$_2$\-Cu$_2$(CN)$_3$.}
\textbf{(A)}~Layers of tilted BEDT-TTF dimers in the $bc$-plane are separated in $a^*$-direction [$a^* \perp (bc)$] by [Cu$_2$(CN)$_3$]$_{\infty}^-$ anion sheets.
\textbf{(B)}~The dimers are internally coupled by $t_d$ and arranged on a slightly distorted triangular lattice: the interdimer transfer integrals $t$ and $t^{\prime}$ define the degree of frustration, $t^{\prime}/t \approx 0.83$.
\textbf{(C)}~Sketch of a valence-bond-solid state on an $S=\frac{1}{2}$ triangular lattice with
spin singlets denoted in blue. Domain walls (top right), topological defects and monomers are expected in real materials. The orange arrows represent unpaired spins due to random pinning of local moments; vacancies are re\-pre\-sented by a green circle.
}
\end{figure}

\clearpage
\begin{figure}
\includegraphics[width=\textwidth]{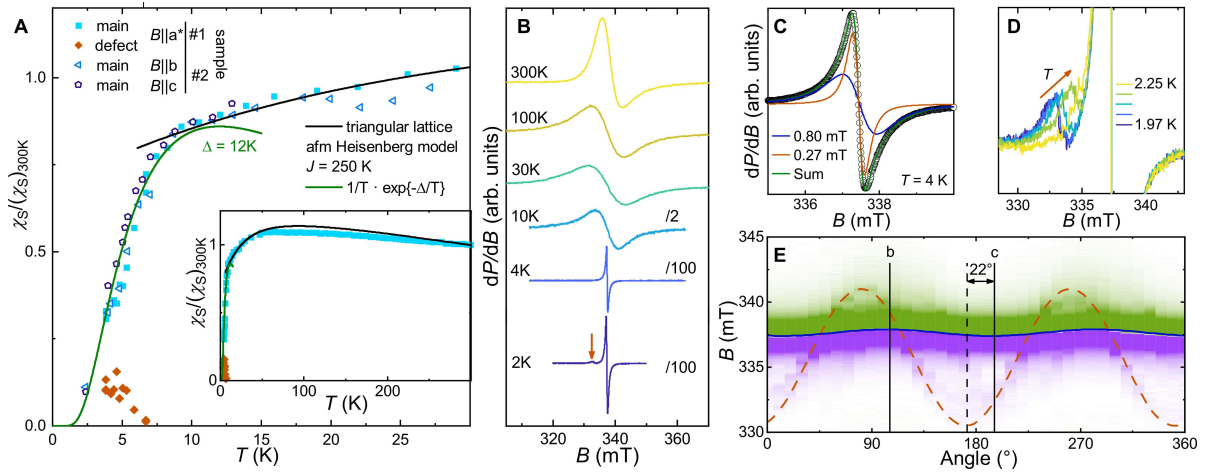}
\caption{
\textbf{X-band ESR results of $\kappa$-(BEDT\--TTF)$_2$\-Cu$_2$(CN)$_3$.}
\textbf{(A)}~Temperature dependence of the normalized spin susceptibility $\chi_S$ measured on various samples and along different directions. At elevated  temperatures, $\chi_S(T)$ is described by an antiferromagnetic Heisenberg model on a triangular lattice with $J=250\,$K (black line). Below the renowned anomaly at $T^*=6\,$K, an exponential decay of the main signal evidences the opening of a spin gap $\Delta = 12\,$K (green line). The orange diamonds correspond to the defect spins, which become obvious for $T<T^*$.
\textbf{(B)}~Temperature evolution of the X-band spectra with the magnetic field $B \parallel c$.
The signal at $T=10\,$K and lower is divided by 2, respectively 100 to account for the increasing peak in ${\rm d}P/{\rm d}B$ as the line sharpens.
\textbf{(C)}~Below $T^*$  an additional narrow component appears requiring
a second Lorentz function to fit the spectra satisfactorily. As an example, the $4\,$K data are shown with the respective decomposition.
\textbf{(D)}~Upon cooling below $2.5\,$K this new signal separates and shifts to lower resonance fields, which we assign to defect spins not involved in the singlet formation.
\textbf{(E)}~Anisotropy of the ESR resonances of $\kappa$-(BEDT\--TTF)$_2$\-Cu$_2$(CN)$_3$. The main signal (solid blue line) is identified by the zero-crossing of ${\rm d}P/{\rm d}B$ from positive (green) to negative (violet), and shows a small angular variation of $0.3\,$mT when measured within the $bc$-plane at $T=2\,$K. In addition, we observe the signal of the defect spins (dashed orange line) with a huge anisotropy of $10\,$mT offset by an angle of $22^\circ$, caused bz dipolar interaction to local moments, possibly Cu$^{2+}$.
}
\end{figure}

\clearpage
\begin{figure}
\includegraphics[width=\textwidth]{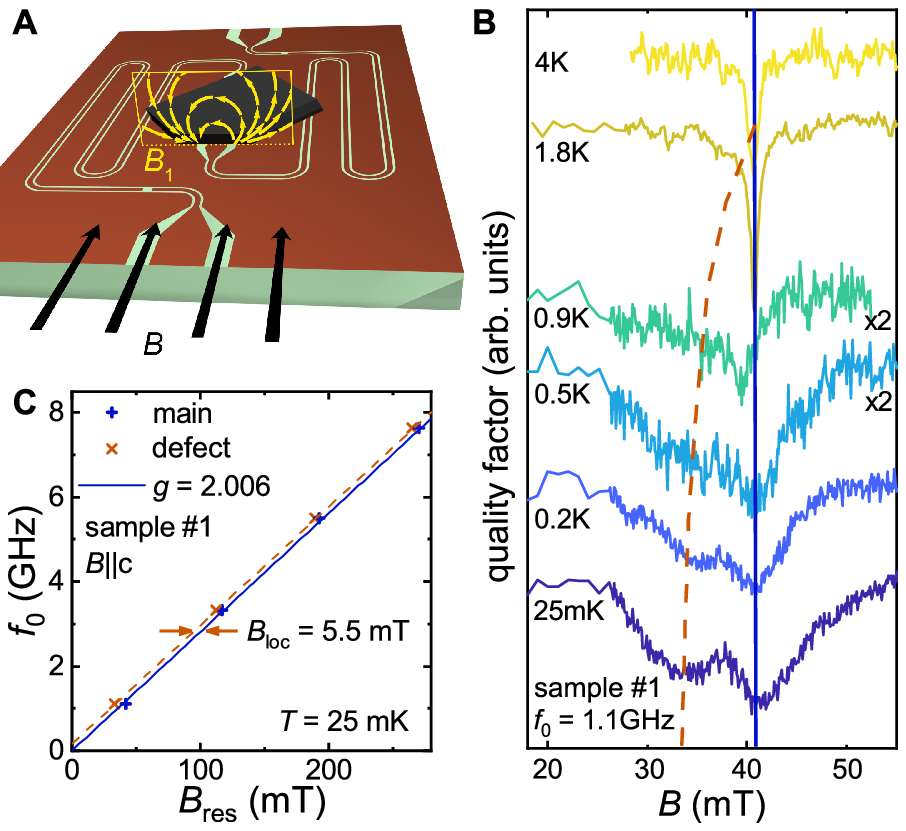}
\caption{
\textbf{Low-temperature broad-band ESR features of $\kappa$-(BEDT\--TTF)$_2$\-Cu$_2$(CN)$_3$.}
\textbf{(A)}~Coplanar waveguide resonator with mounted sample. The direction of the static magnetic field $B$ is shown in black. The microwave magnetic field $B_1$ (yellow) is subject to absorption by electron spin resonance within the sample.
\textbf{(B)}~$B\parallel c$ ESR spectra at $1.1\,$GHz for different temperatures down to $25\,$mK.
The solid blue line corresponds to the constant main signal $B_\textrm{main}$, while
the dashed orange line indicates the evolution of the local-moment signal.
\textbf{(C)}~Low-temperature peak positions for varying resonance frequencies and fields.
}
\end{figure}

\end{document}



\maketitle
\tableofcontents

\clearpage
\section{Samples}
\begin{figure}[hb]
  \centering
  \includegraphics[width=0.35\textwidth]{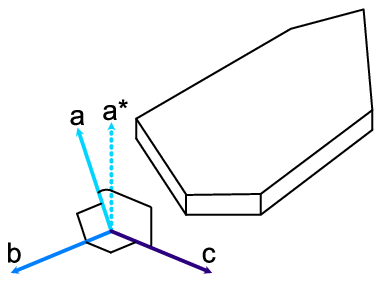}
  \caption{Crystal axes of \CuCN, which possesses the monoclinic space group P2$_1$/c. The room-temperature lattice parameters are $a=\SI{16.0920}{\angstrom}$, $b=\SI{8.5813}{\angstrom}$, $c=\SI{13.3904}{\angstrom}$, $\beta=\SI{113.381}{\degree}$ \cite{Pinteric14}. The $b$ and $c$-directions span the plane of the two-dimensional spin system. $a^*\perp(bc)$ is the out of plane direction and coincident with the thickness of the crystal platelet. }
  \label{fig:coordinats}
\end{figure}

The experiments were performed on \CuCN\ single crystals grown by electro-crystallization methods \cite{Geiser91} at four different laboratories:
\begin{itemize}
\item[\#\,1] Institute of Problems of Chemical Physics, Russian Academy of Sciences, Cherno\-go\-lovka, Russia 
\item[\#\,2] Department of Applied Physics, University of Tokyo, Tokyo, Japan 
\item[\#\,3] 1.~Physikalisches Institut, Universität Stuttgart, Stuttgart, Germany. 
\item[\#\,4] Material Science Division, Argonne National Laboratory, Argonne, Illinois, U.S.A. 
\end{itemize}
The single crystals have plate-like shape with a typical size of $\SI{1}{mm}\times\SI{1.5}{mm}\times\SI{100}{\micro\meter}$  as shown in Fig.~\ref{fig:coordinats}.
They were measured as grown without further treatment.

Crystals from each batch were characterized by dc resistivity measurements as a function of temperature.
The polarized infrared reflectivity was measured at room temperature in order to select specimens of the correct phase and determine their crystallographic orientation.
Precise alignment (within \SI{5}{\degree}) of the crystals for the ESR experiments was achieved in-situ at the X-band ESR spectrometer utilizing a goniometer.
The temperature-dependent ESR parameters show a close resemblance among the samples for $T>\SI{4}{K}$.
For lower temperatures, significant differences were observed, closely related to the emerging defect signal.
Therefore, angle-dependent ESR measurements at $T=\SI{2}{K}$ have been performed to scrutinize the sample-to-sample variation of the low temperature properties.

\clearpage

\section{X-band and W-band ESR spectroscopy}

Conventional electron-spin-resonance (ESR) spectroscopy was performed on a Bruker EMXplus continuous wave spectrometer operating in the X-band, i.e.\ $f_0\approx\SI{9.5}{GHz}$ and $B\approx\SI{340}{mT}$, employing a standard rectangular TE$_{102}$ cavity (Bruker ER 4102ST). It is equipped with an Oxford Instruments ESR900 continuous flow helium cryostat to conduct temperature-dependent experiments in the range $T=4-\SI{300}{K}$. To reach lower temperatures down to $T=\SI{1.9}{K}$, a pumped cryostat (Oxford Instruments ESR910) was used. The spectrometer allows us to measure the anisotropic ESR signal of single crystals with the help of a programmable goniometer (Bruker ER 218PG1). Measurements at higher frequency were performed using a Bruker ELEXYS E680 W-band spectrometer operating at $f_0\approx\SI{95}{GHz}$.

\subsection{Sample characterization}

\begin{figure}
  \centering
  \includegraphics[width=0.7\linewidth]{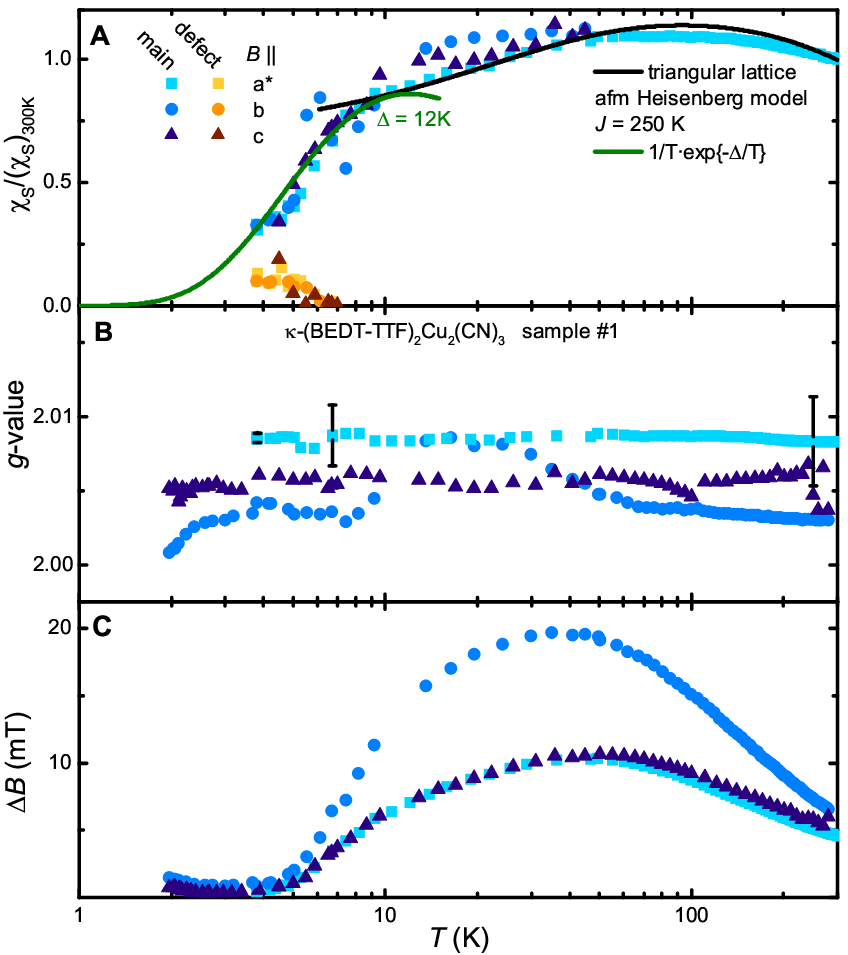}
  \caption{
  Temperature dependence of the (\textbf{A})~spin susceptibility $\chi_S$, (\textbf{B})~$g$-value, and (\textbf{C})~line width $\Delta B$ obtained from fits to the ESR spectra measured in the three directions ${B} || a^*, b, c$ for \CuCN\ sample \#\,1.
  Measurements have been performed in the X-band (\SI{9.5}{GHz}).
  The spin susceptibility is described by an antiferromagnetic Heisenberg model on a triangular lattice at elevated temperatures. Below $T^*=\SI{6}{K}$ an exponential decay of the main signal evidences the opening of a spin gap
  $\Delta = \SI{12}{K}$. Below $T^*$ we notice the simultaneous appearance of the defect component:
  The orange symbols corresponds to the defect spins.
  The $g$-value does not exhibit any change with temperature within the experimental uncertainty estimated as 5\% of the line width.
  The line width $\Delta B(T)$ is largest along the $b$-direction. Starting from room temperature, it first increases similar to other \ET\ compounds \cite{Nakamura94}, reaching a maximum around \SI{40}{K} along all three directions. Below the line starts to narrow; a drastic decrease of the line width sets in below \SI{10}{K} and a minimum is reached around $T_\text{loc}$.
  }
  \label{fig:Xband}
\end{figure}

X-band ESR measurements as a function of temperature for the magnetic field
oriented along the three crystallographic axes $a^*$, $b$, $c$
provide a first characterization of the magnetic properties.
The observed parameters coincide well with previous results \cite{Nakamura94,Komatsu96,Padmalekha15}.

In Fig.~\ref{fig:Xband}A we plot the temperature-dependent spin susceptibility $\chi_S(T)$.
In the high-$T$ regime, $\chi_S(T)$ can be described by the antiferromagnetic Heisenberg model on a triangular lattice.
The black curve shown Fig.~\ref{fig:Xband}A is the [7/7] Pad{\'e} approximant of the high-temperature series expansion from \cite{Elstner93} with $J=\SI{250}{K}$ in accord with the literature \cite{Shimizu03}. 

Below $T^*=\SI{6}{K}$ the formation of a valence-bond-solid spin-singlet state results in an exponential decrease of $\chi_S(T)$.
For details of the behavior and fits by appropriate models see Section~\ref{sec:spingap}.
The observation is confirmed by crystals of different origin. As a matter of fact, first indications of a spin gap can be found in previous ESR data of Komatsu {\it et al.} \cite{Komatsu96}; a pronounced Curie tail below $T^*$, however, concealed the opening of the spin gap. The susceptibility data from SQUID measurements \cite{Shimizu03} also show a drop by 30\%\ below $T^*=\SI{6}{K}$, which is in accord with our results. Similar conclusions can be drawn from measurements of the NMR spin-lattice relaxation rate, that exhibits a small but noticeable drop in $1/(T_1T)$ seen both $^1$H and $^{13}$C below $T^*$ \cite{Shimizu03,Shimizu06}, before a strong increase is observed that we assign to defects in line with previous NMR work on related \ET\ \cite{Pustogow19}.

Simultaneously, an additional narrow component appears in the spectrum at the same resonance field. 
The intensity, extracted from a fit by two Lorentzians below \SI{6}{K}, is plotted in Fig.~\ref{fig:Xband}A by the orange diamonds.
We suggest that this signal originates from unpaired defect spins in the valence-bond-solid state.

Cooling further leads to a splitting of the spectra below $T_\text{loc}\approx \SI{2.5}{K}$. Part of the defect signal starts to shift away from $B_{\text{main}}\simeq 337$~mT. The intensity of this emerging signal is strongly sample-dependent.
At $T = \SI{2}{K}$ its intensity much more pronounced for sample \#\,3 as compared to samples \#\,1, \#\,2 and \#\,4. Accordingly, we can trace well separated peaks up to $T_\text{loc}$ for sample \#\,3, while for sample \#\,2, for instance, it cannot be identified any more due to the smaller intensity.
For crystals from a certain laboratory and within a batch, the sample-to-sample variation of the defect signal intensity is smaller (less than a factor of 2). From this pronounced sample dependence we conclude that intrinsic disorder is of superior importance for the low-temperature properties.

It is interesting to compare our findings with the sample-to-sample variation observed for the so-called 6~K-anomaly at $T^*$ observed by thermal expansion measurements \cite{Manna18}. While the feature is always present and does not shift in temperature, the size of the anomaly varies significantly indicating a broadening of the transition, like also seen for spin-Peierls transitions subject to disorder.

For both ESR-bands the $g$-value of the intrinsic spin contribution is temperature-independent for all orientations:
within the plane $g_b=2.004$ and $g_c=2.006$, while out-of-plane $g_{a^*}=2.008$. 
The results agree with literature values \cite{Komatsu96,Padmalekha15}.

\begin{figure}
  \centering
  \includegraphics[width=0.7\linewidth]{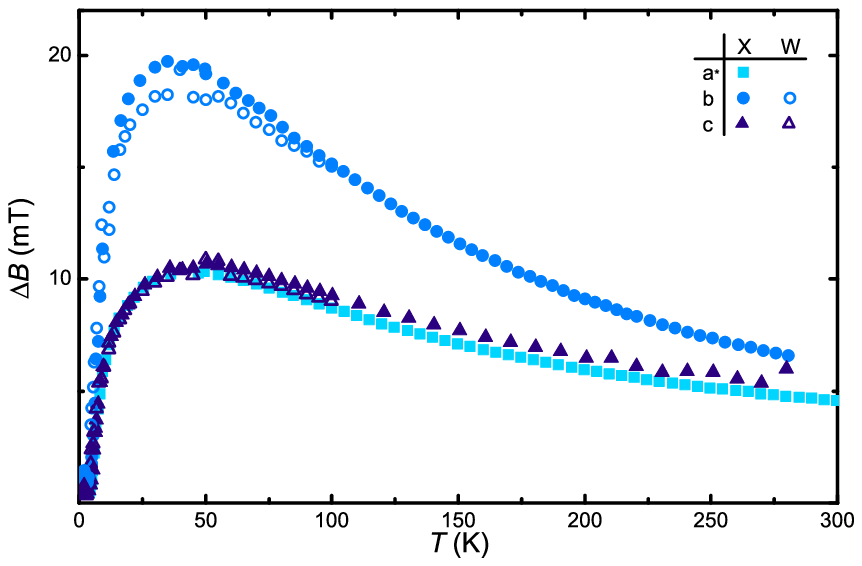}
  \caption{
  Temperature dependence of the line width $\Delta B$ at different external magnetic field.
  X-band data (closed symbols) measured at $B\approx\SI{340}{mT}$ and W-band data (open symbols) measured at 10 times higher field $B\approx\SI{3.4}{T}$ show almost identical line width.
  }
  \label{fig:Wband}
\end{figure}

\begin{figure}
  \centering
  \includegraphics[width=0.5\linewidth]{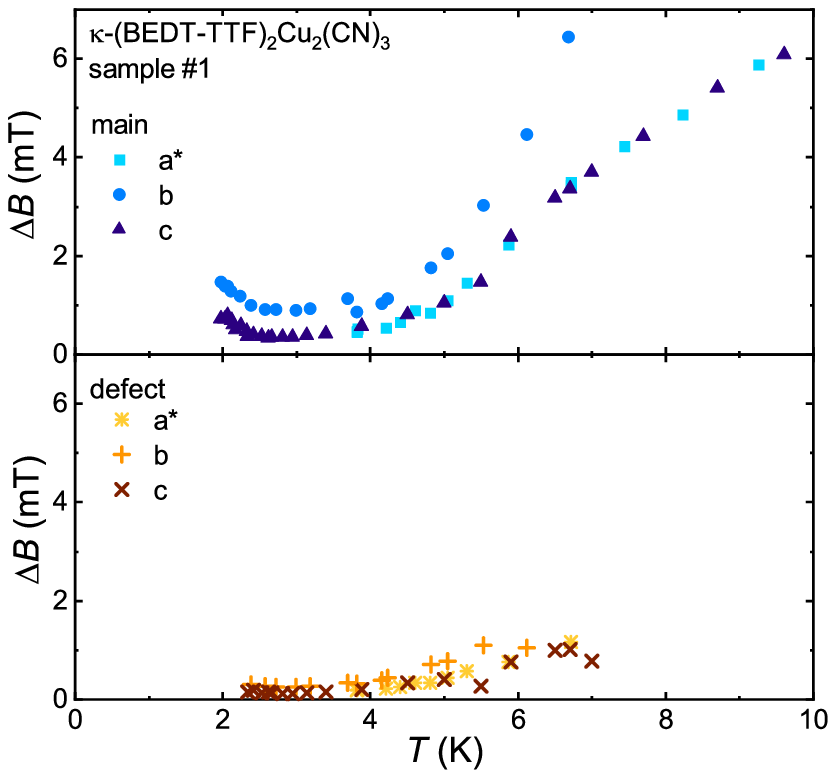}
  \caption{
  Magnification of the low-temperature behavior of the line width for the main signal as well as the defect line of \CuCN\ sample \#\,1  measured along all three directions using X-band spectroscopy. While all lines become continuously sharper upon cooling down to $T=\SI{3}{K}$, a pronounced broadening is observed below \SI{2.5}{K} that is also seen in Fig.~3B.}
  \label{fig:lowTlinewidth}
\end{figure}

Cooling below room temperature, the line width increases,
reaches a broad maximum around \SI{40}{K}, before it is strongly reduced (Fig.~\ref{fig:Xband}C).
$\Delta B$ is about twice as large along the $b$-axis compared to the other directions. 
Measurements in the W-band yield identical line widths as shown in Fig. \ref{fig:Wband}, which rules out inhomogeneous broadening.
Below $T=\SI{2.5}{K}$ the line width starts to increase again, as is shown in Fig.~\ref{fig:lowTlinewidth} on a magnified scale.
As the intrinsic signal is almost frozen out at this temperature the increase can be attributed to the defect component.
This behavior coincides with the incipient separation of the defect contribution from $B_{\text{main}}$.
The minimum of $\Delta B(T)$ around $T \approx \SI{3}{K}$ shown in Fig.~\ref{fig:lowTlinewidth} coincides with features that can be identified in specific heat data at the same temperature \cite{Yamashita08,Manna10}.

Because of the large exchange coupling of $J=\SI{250}{K}$ an exchange narrowed ESR absorption is expected. The line width can then be approximated as \cite{Yamada96}
\begin{equation}
  \Delta B \simeq \frac{\hbar^2}{g\mu_\text{B}|k_\text{B}J|}M_2(J/T) \quad ,
\end{equation}
where $M_2(J/T)$ is the second moment corresponding to the relevant perturbation.
In the high temperature limit $T>J$, the contribution of different processes can be estimated \cite{Yamada96,Taniguchi97}:
dipole-dipole interaction $M_2^{DD}(0)$, anisotropic exchange $M_2^{AE}(0)$ and antisymmetric -- Dzyaloshinsky-Moriya (DM) -- exchange $M_2^{AE}(0)$.
Using the values for the exchange coupling $J$, the DM vector $\vek{D}$ and the anisotropic exchange tensor $\vek{\Gamma}$ from \cite{Winter17}, the individual contributions can be calculated.
This reveals the DM exchange as the dominant contribution to the line width with $\Delta B^{DM}\approx 10-\SI{15}{mT}$, and negligible contribution of dipole-dipole interaction and anisotropic exchange, both in the range of $10^{-3}\,$mT.

Due to the larger conductivity at high temperatures, the increased spin diffusion reduces the line width according to Elliott's relation \cite{Elliott54}. The value of $\Delta B$ as well as its anisotropy around the broad maximum is in the range of the calculated high temperature limit. A decrease of the DM contribution to the line width upon cooling is expected \cite{Yamada96}. The increase below \SI{2.5}{K} can be attributed to an increasing contribution of dipole-dipole interaction of the emerging defect spins with local impurities that are increasingly polarized at lower temperatures.

\clearpage

\subsection{Models for the Spin Gap}
\label{sec:spingap}
\begin{figure}[ht]
  \centering
  \includegraphics[width=0.5\linewidth]{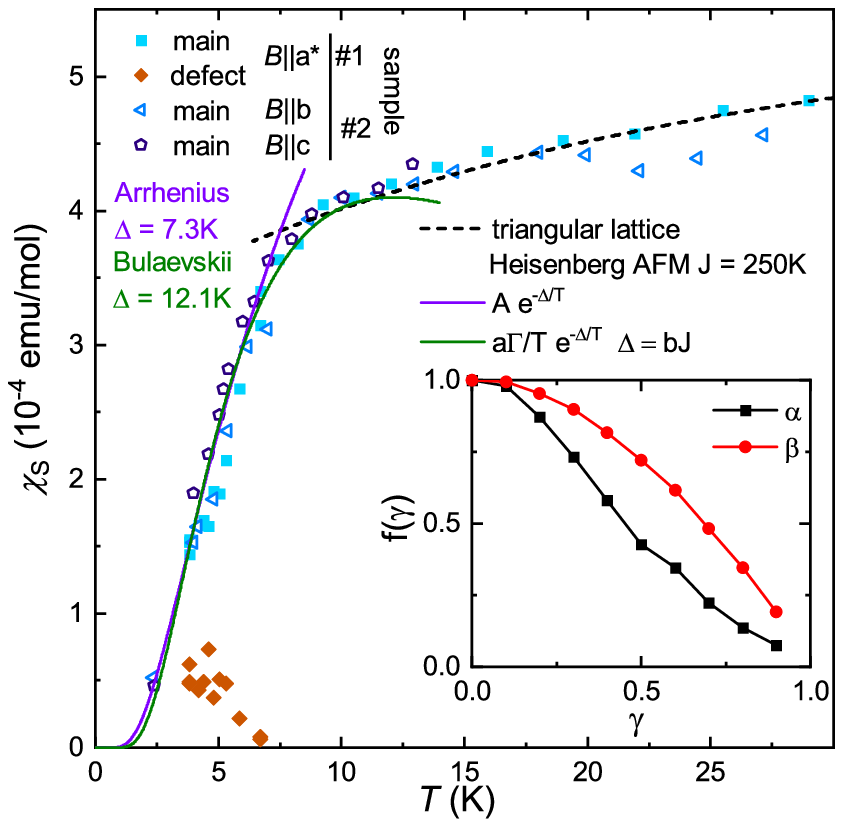}
  \caption{
  Absolute values of the spin susceptibility $\chi_S$ of \CuCN\ below $T=\SI{30}{K}$.
  Two slightly different approaches are compared to describe the exponential decrease due to the opening of a spin gap: the simple activated behavior follows Arrhenius law (violet), and the Bulaevskii model assumes an alternating spin chain (green). The inset shows how the parameters of Bulaevskii's model $\alpha$ and $\beta$ depend on the alternation parameter $\gamma$ \cite{Bulaevskii69}; see text for more details.
  }
  \label{fig:spin_gap}
\end{figure}

At elevated temperatures, $T>T^*$,  $\chi_S(T)$ behaves like an $S=\frac{1}{2}$ Heisenberg antiferromagnet on a triangular lattice \cite{Elstner93}; the extracted exchange interaction $J=\SI{250}{K}$ is rather strong, as reported previously \cite{Shimizu03}.
Below $T^*$ the spin susceptibility $\chi_S(T)$ sharply decreases below the predictions of the Heisenberg model
indicating the formation of a non-magnetic ground state. In other words, the spins on two adjacent BEDT-TTF dimers pair to an $S=0$ state with only singlet-triplet excitations possible, separated by a spin gap.
In order to estimate the energy gap, we fit the susceptibility data by two models as shown in Fig.~\ref{fig:spin_gap}:
\begin{align}
\text{activated behavior (Arrhenius law):}\qquad\chi_S &= A\exp\left\{-\frac{\Delta}{T}\right\}\\
\text{alternating spin chain (Bulaevskii):}\qquad\chi_S &= \alpha\Gamma\exp\left\{-\frac{\beta|J_1|}{T}\right\}
\end{align}
In the simplest case of a thermally activated behavior described by the Boltzmann distribution, we obtain a spin gap $\Delta=\SI{7.3}{K}$ with a prefactor $A=\SI{1.02e-3}{emu\per\mole}$.

The Bulaevskii model \cite{Bulaevskii69} describes the temperature-dependent susceptibility of an alternating spin chain, where the exchange couplings $J_1$ and $J_2$ alternate along the chain. 
The model has been extended to include lattice dynamics to account for temperature-dependent spin dimerization from a uniform spin-chain to an alternating one to model the spin-Peierls phase transition \cite{Pytte74,Bray75,Johnston00}, for instance. The dimerization strength $J_2/J_1=\gamma$ can be related to the exchange coupling $J$ of the uniform spin chain above the transition temperature with the relation $J_{1,2}=J(1\pm\delta)$ with $\delta=(1-\gamma)/(1+\gamma)$. The prefactor $\Gamma=N_A g^2\mu_\text{B}^2/k_\text{B}=\SI{15012}{emu\,\kelvin\per\mole}$ related to the Curie constant is used to obtain the molar susceptibility. 

Strong in-plane anisotropy of thermal expansion causes our two-dimensional spin system to develop one-dimensional character.
We thus apply the approach to \CuCN\ as it is the simplest model to capture the relevant spin dimerisation.
Using the exchange coupling $J=\SI{250}{K}$, the fit shown in Fig.~\ref{fig:spin_gap} leads to $\alpha=9.0\times 10^{-3}$ and $\beta=0.047$. This implies a spin gap of $\Delta=\SI{12.1}{K}$. From the parameters $\alpha$ and $\beta$ one can estimate an alternation parameter of $\gamma=0.97$ according to the values from \cite{Bulaevskii69} as shown in the inset in Fig. \ref{fig:spin_gap}; in other words the
valence bonds in Fig.~1C enhance their coupling by $\leq 2\%$. According to mean-field calculations the dimerization temperature is related to the gap via the BCS-like relation $2\Delta(0)/T^*=3.53$ leading to a $T^*=\SI{6.9}{K}$ in accord with the transition temperature observed in $\chi_S(T)$ and the sharp feature in the lattice expansion seen by thermal expansion experiments \cite{Manna10}.
We thus propose a valence-bond-solid transition at $T^*=\SI{6}{K}$ coupled to the lattice similar to the case of spin-Peierls transitions in one-dimensional spin chains. The anisotropy of the thermal-expansion anomaly indicates that the dimerization occurs preferably along the $c$-axis.

\clearpage

\subsection{\ce{Cu^2+} impurities}
\label{sec:Cu2+}
\begin{figure}[hb]
  \centering
  \includegraphics[width=0.7\linewidth]{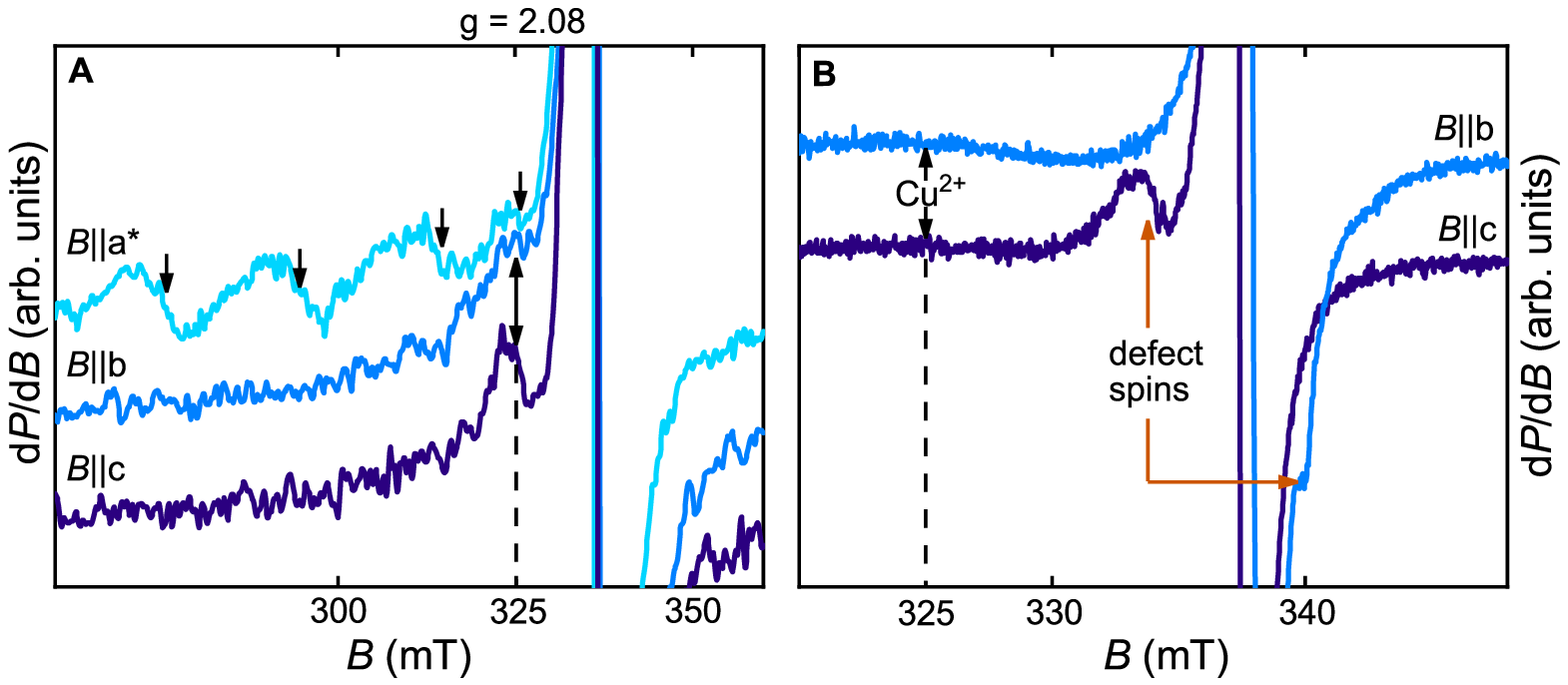}
  \caption{
  Comparison of the \ce{Cu^2+} signal (\textbf{A}) present in all studied samples of \CuCN\  and the observed (BEDT-TTF)$_2$ defect signal (\textbf{B}) at $T=\SI{4}{K}$.
  The \ce{Cu^2+} signal shows almost no anisotropy within the $bc$-plane with $g_{||c} = 2.08$.
  The anisotropy of the low temperature defect signal is large as it moves from the low-field side of the main peak for ${B}\parallel c$ to the high-field side for ${B}\parallel b$.
  }
  \label{fig:cu2+}
\end{figure}

The spectra displayed in Fig.~\ref{fig:cu2+} evidence the presence of $S=1/2$  \ce{Cu^2+} impurities
in \CuCN\ in accord with previous reports \cite{Komatsu96,Padmalekha15}.
In panel (A) the characteristic structure of the \ce{Cu^2+} ESR signal can be observed. Within the $bc$-plane the lines collapse into a single line and show almost no anisotropy. In the $B\parallel a^*$-direction four equally separated lines are visible;
i.e.\ perpendicular to the layers.
The splitting into four lines is due to anisotropic hyperfine interaction with the copper nuclei with spin $I=3/2$. The observed $g$-values of $g_{b,c}=2.08$ and $g_{a^*}=2.22$ fit well to the values obtained from molecular orbital calculations in Ref.~\cite{Komatsu96}.
These characteristic peaks appear only below $T=\SI{10}{K}$. Their intensity is strongly sample-dependent ranging from around \SI{50}{ppm} to \SI{1350}{ppm} for the sample shown in Fig.~\ref{fig:cu2+}, but we could not find crystals without indications of \ce{Cu^2+}.

Panel (B) shows the anisotropic behavior of the signal of the (BEDT-TTF)$_2$ defect spins,
which differs fundamentally from the \ce{Cu^2+} signal, the latter being much broader. With ${B}\parallel c$, it appears at lower field than the main signal. Upon rotating the field to ${B}\parallel b$ it moves through the main signal to higher field, as discussed in more detail in Section~\ref{sec:anisotropy}. The different anisotropy and line width rule out \ce{Cu^2+} impurities as the source of the observed defect signal; we will discuss on dipolar interaction among these two types of magnetic moments in section~\ref{sec:couplingCu2+}.
Furthermore, the offset to lower fields of the \ce{Cu^2+}-lines is much larger than the
additional signal from the defect spins.

\clearpage

\subsection{Structural Twinning}
\label{sec:twinning}
\begin{figure}[hb]
  \centering
  \includegraphics[width=1\linewidth]{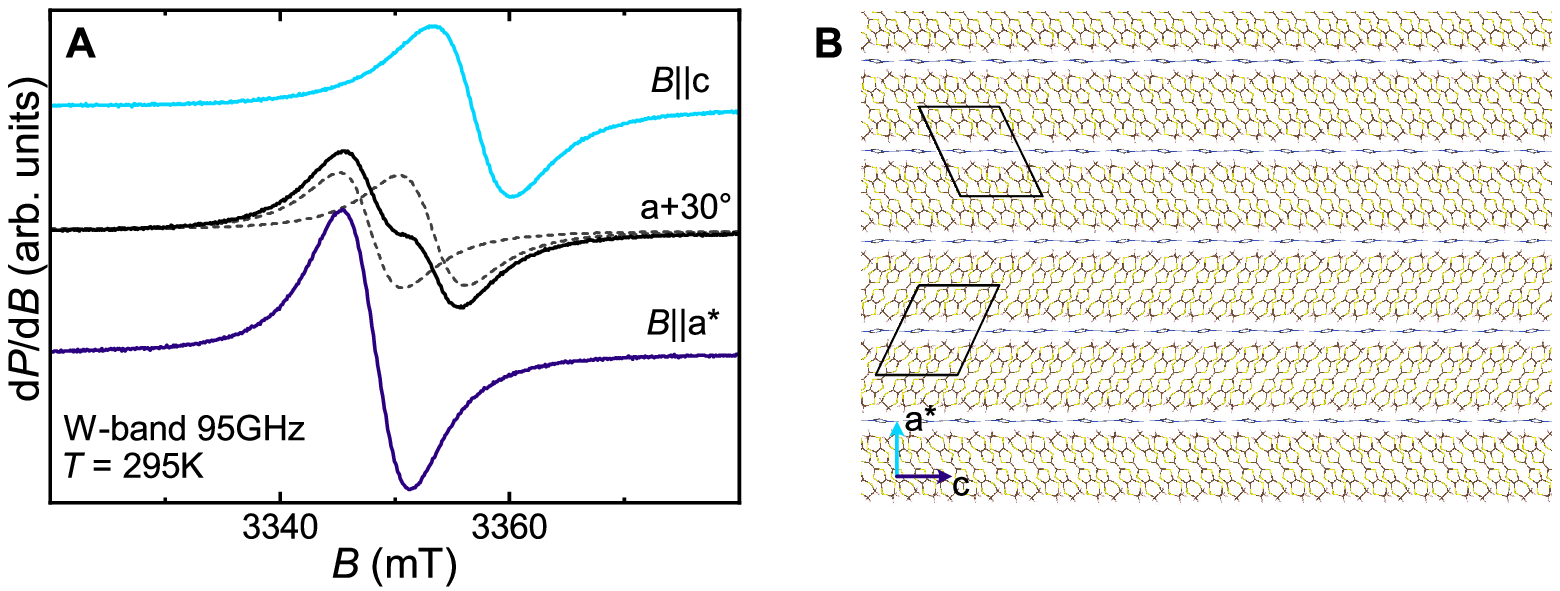}
  \caption{ESR anisotropy discloses crystal twinning. (\textbf{A})~Angular dependence of the W-band signal when rotating the magnetic field $B$
  within the $a^*c$-plane.
  Most samples exhibit a splitting at intermediate angles, inferring that the crystals are twinned as sketched in panel (\textbf{B}). The two lines always collapse in the $a^*$ and $c$-directions.}
  \label{fig:twinning}
\end{figure}

Many samples show evidence of twinning.
This is particularly clear in the angular dependence of the room-temperature W-band spectra
displayed in Fig.\ \ref{fig:twinning}A.
While along the $a^*$ and $c$-directions, we always find a single Lorentzian line shape, for intermediate angles the spectra split in two lines; here illustrated for the example of $a^*+30^{\circ}$.
Such behavior is known \cite{Antal09} from those $\kappa$-phase compounds with BEDT-TTF molecules arranged in a herring bone fashion, such as \CuCl. In those cases the orientation of the organic molecules is tilted in alternating direction when looking in the direction perpendicular to the layers.
The crystallographically inequivalent adjacent layers lead to two signals which collapse along and perpendicular to the layers.

For the monoclinic \CuCN\ crystals, however, such a behavior is  not expected. Hence, the observation of  a splitting in the ESR spectra indicates twinning of the crystals as sketched in panel B.
Domains of ``left-tilted'' molecules are separated by stacking faults from domains of ``right-tilted'' molecules.
These findings are robust: from a large number of crystals of different sources and batches, basically all specimens show hints of twinning. While the angular dependence is always identical, the lines acquire different intensities, reflecting the ratio of ``left''- to ``right-tilted'' domains.
There are few exceptions, like for sample \#\,1 displayed in Fig.~2E of the main paper, which possess more or less a single line, i.e.\ are single-domain with very weak twinning.

\clearpage

\subsection{Low-temperature anisotropy of the defect signal}
\label{sec:anisotropy}
\begin{figure}[hb]
  \centering
  \includegraphics[width=0.9\linewidth]{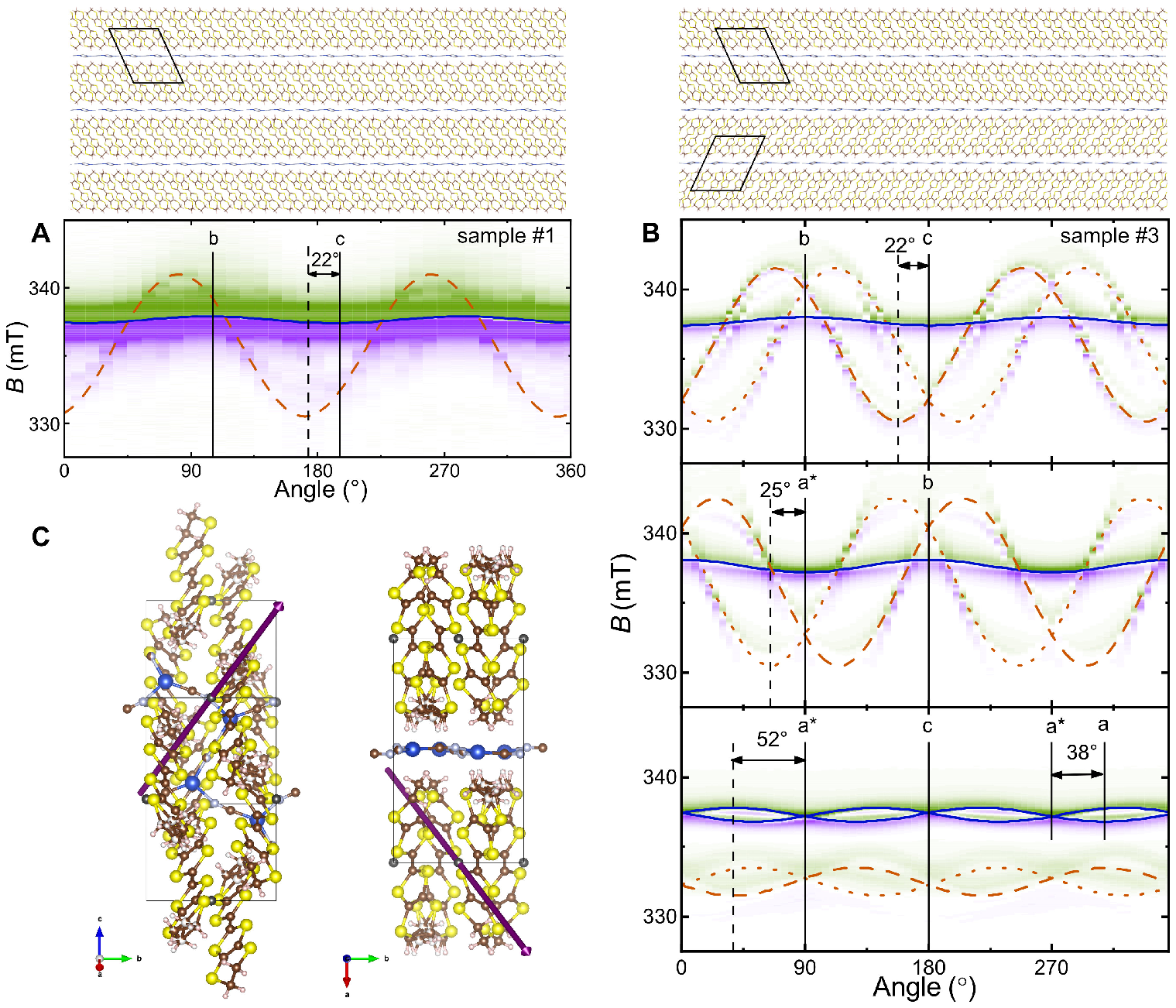}
  \caption{
  Anisotropy of the X-band ESR resonance field of \CuCN\ measured at $T=\SI{2}{K}$ for the single-domain sample \#\,1 (\textbf{A}), and the twinned crystal \#\,3 (\textbf{B}).
  Besides the main signal from the BEDT-TTF radical  (solid line), an additional  signal appears at low-temperatures (dashed and dotted lines).
  We relate the splitting of the main signal (for rotation within the $a*c$ plane) as well as the doubling of the defect signal to twinning in the crystal as illustrated in the top frames.
  (\textbf{C}) The orientation of the dipole interaction responsible for the shifted defect signal is shown with the purple arrows within the crystal structure viewed along the $c$ and $a*$-direction.
  }
  \label{fig:lowt_anisotropy}
\end{figure}

In Fig.\ \ref{fig:lowt_anisotropy} the angular dependence of the ESR signals is plotted for samples \#\,1 and \#\,3, as recorded at $T=\SI{2}{K}$.
In panel A we reproduce Fig.~2E of the main paper, where the angular dependence within the $bc$-plane is presented
for sample \#\,1.
In the case of sample \#\,3 [Fig.~\ref{fig:lowt_anisotropy}B] the full anisotropy is shown with the static field rotated within the $bc$-plane (upper), the $a^*c$-plane (middle) and the $a^*b$-plane (lower panel).

The resonance field of the main signal (solid line) varies only slightly upon rotation, well described by an anisotropic $g$-tensor.
Within the principal axes system $a^*, b, c$ the $g$-value is given by
\begin{equation}
  g=\sqrt{g_{a^*}\cos^2\theta+g_b \sin^2\theta\cos^2\phi+g_c\sin^2\theta\sin^2\phi}
\end{equation}
with $\theta$ describing the angle of the external magnetic field with respect to the out-of-plane $a^*$ direction, and $\phi$ the angle within the $bc$-plane.
The resonance field of the main signal is therefore following a $\cos^2$ angular dependence for any of the chosen axes of rotation. 

\begin{figure}
  \centering
  \includegraphics[width=0.9\linewidth]{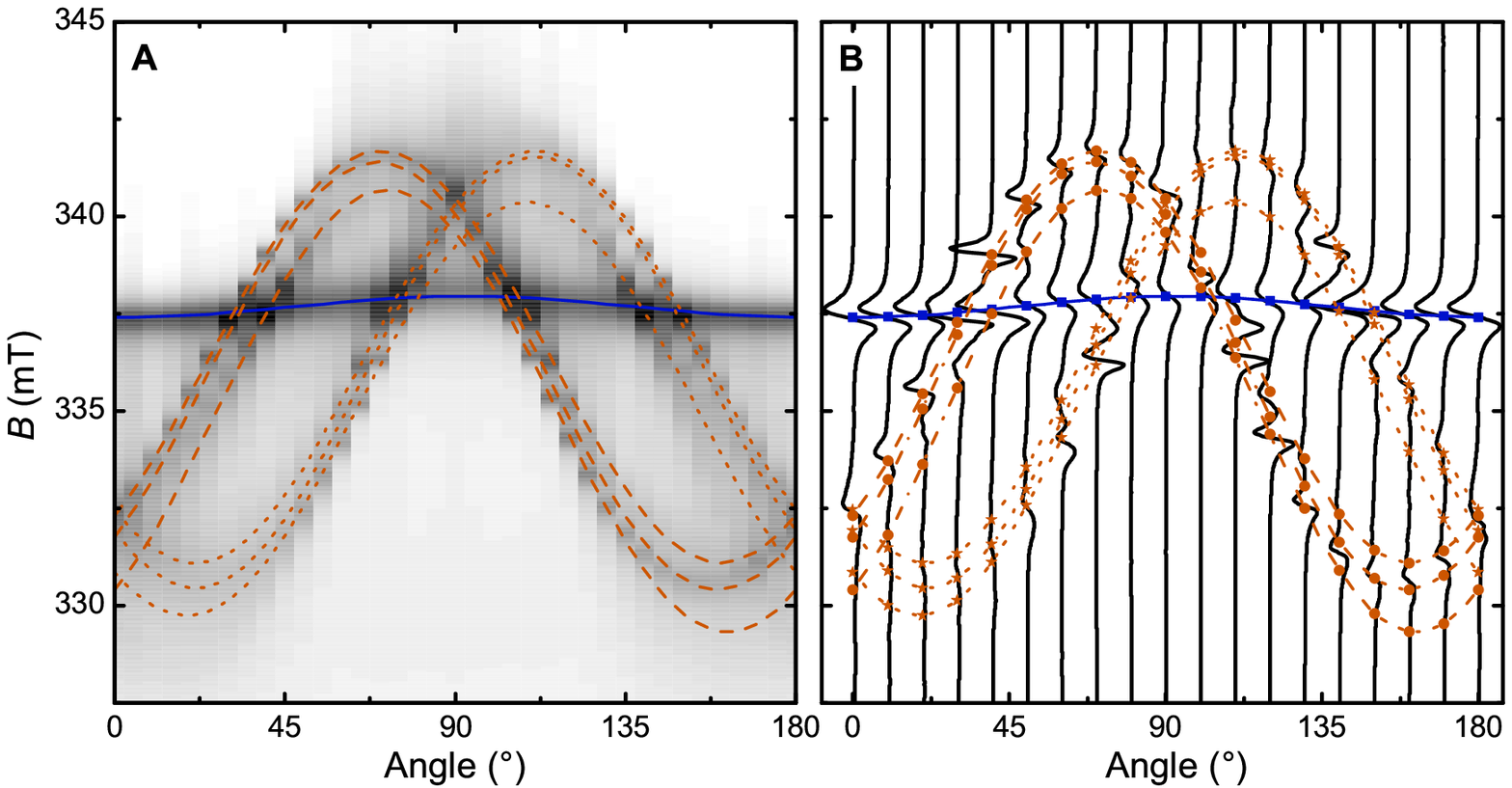}
  \caption{
  Detailed view of the anisotropy in the $bc$-plane for sample \#\,3 at $T=\SI{2}{K}$.
  The two sets of defect peaks in the twinned sample have been fitted with 3 Lorentzians each.
  The resonance fields of the main signal (blue) and defect peaks (orange) are shown on top of the absorption (\textbf{A}) and derivative \textbf{(B)} spectra.
  }
  \label{fig:anisotropy_multiple}
\end{figure}

\begin{figure}
  \centering
  \includegraphics[width=0.45\linewidth]{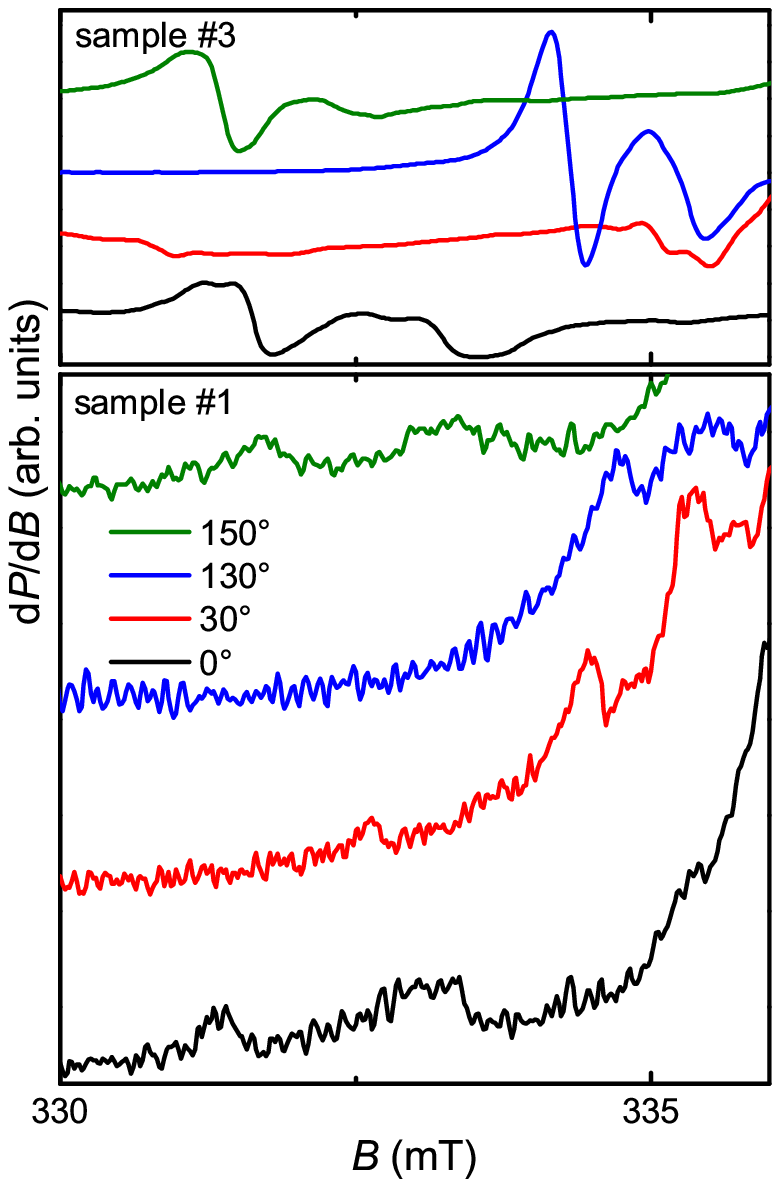}
  \caption{
  Comparison of the components of the defect signal for samples \#\,1 and \#\,3 (same data as in Fig.~\ref{fig:anisotropy_multiple}) at 2 K with the magnetic field within the $bc$-plane.
  }
  \label{fig:defect_components}
\end{figure}

Twinning has to be taken into account to fully describe the anisotropy of the ESR spectra.
On the one hand, sample \#\,1 does not show a splitting of the main signal, which  proves that the crystal is single-domain.
We also observe only one set of 2--3 defect lines in the spectrum shifted by $- 22^{\circ}$, which will be discussed later in relation to Figs.~\ref{fig:defect_components} and \ref{fig:anisotropy_multiple}.
For sample \#\,3, on the other hand, clear signs of twinning are present:
For the $a^*c$-plane a splitting is detected already at room temperature as shown in Fig.~\ref{fig:twinning}.
Here, the minima in resonance field are symmetrically shifted by $\pm 38^{\circ}$ with respect to the $a^*$-direction, which is exactly the tilt angle of the BEDT-TTF molecules and comparable to \CuCl.
This splitting and the two visible defect signals with symmetric anisotropy around the crystallographic axis correspond to the two distinct orientations of the molecules that occur due to twinning as illustrated in the upper frames of Fig.~\ref{fig:lowt_anisotropy}: sample \#\,1 [panel A] consists of a single domain, while sample \#\,3 [panel B] is twinned and contains domains of both orientations.

The behavior of the defect signals is even more surprising because it exhibits a pronounced anisotropy distinct from the main line.
The main line has its maximum along the crystallographic $b$-direction whereas the maxima of the defect signals are shifted by an angle of $\pm 22^\circ$ in the $bc$-plane.
Similar shifts can be identified for the other directions: $\pm 25^\circ$ in the $a^*b$-plane and $\pm 52^\circ$ in the $a^*c$-plane.

From Fig. \ref{fig:lowt_anisotropy}A and B it becomes obvious that the anisotropy of the defect line is much larger (\SI{10}{mT}) compared to the main signal, in particular within the $bc$ and $a^*c$-planes.
The angular shift between the main signal and the defect signal within the $bc$-plane is exactly the same \SI{22}{\degree}  for both samples shown.
The angular dependence of the resonance field $B_\text{res}$ is alike in the different samples despite the very different intensities of the defect contributions.
The resonance shift of the defect signal can be described by a local dipole field $B_\text{loc}$ caused by a localized magnetic moment that is about 6--7 \AA\ distant (for instance \ce{Cu^2+} impurities) acting on the defect spins.
\begin{equation}
  B_\text{loc}(\alpha) = -\frac{\mu_0\mu_\text{B}g S}{4\pi r^3}\left( 3\cos^2(\alpha) - 1 \right)
\label{Eq-dipolar-coupling}
\end{equation}
Its magnitude depends on the angle $\alpha$ between the external magnetic field $B$ and the vector connecting the spin defect and the impurity responsible for the local field indicated in Fig. \ref{fig:lowt_anisotropy}C within the crystal structure of \CuCN.

We conclude the angular dependence of the local fields to be robust against different amounts of disorder in the spin system, albeit its intensity varies from crystal to crystal. 
We observed no significant changes upon different cooling cycles.

A close look to the spectra reveals that the defect signal is comprised of at least three peaks. We fitted three peaks for the strongly twinned sample \#\,3 which exhibits significantly more defects than sample \#\,1. Results are shown in Fig. \ref{fig:anisotropy_multiple}. Even though the intensity of the defect signals is much smaller for sample \#\,1 multiple components of the defect signal are also visible as shown in Fig. \ref{fig:defect_components}. Taking into account a small offset of $5-10^\circ$ between the measurements due to mounting uncertainties they exhibit the same angle dependence as for sample \#\,3.

\clearpage

\subsection{Possible coupling of the defect spins to the \ce{Cu^2+} impurities}
\label{sec:couplingCu2+}
\begin{figure}[hb]
  \centering
  \includegraphics[width=0.7\linewidth]{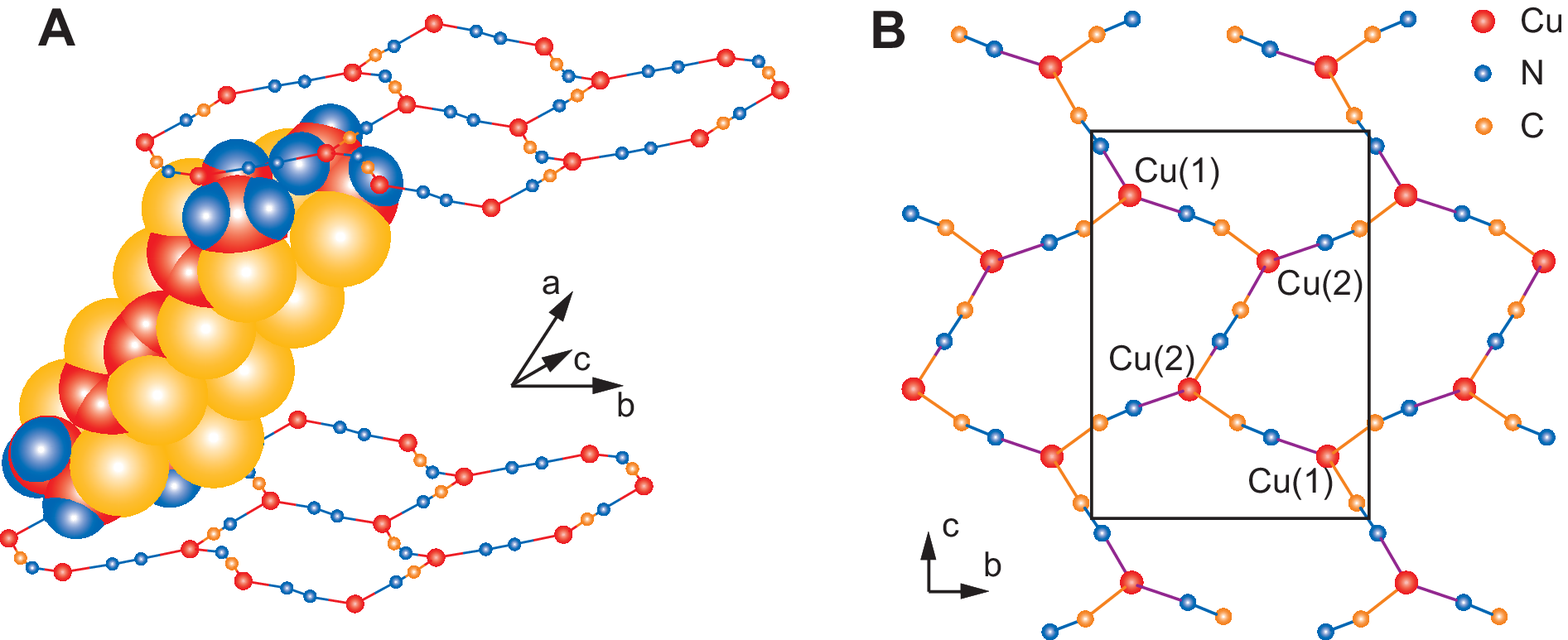}
  \caption{Illustration of the  \CuCN\ crystal structure \cite{Dressel16}.
  (\textbf{A}) The BEDT-TTF layers are sandwiched between polymerized Cu$_2$(CN)$_3^-$ anion sheets, which extend along the $bc$-plane (\textbf{B}).
  }
  \label{fig:structure2}
\end{figure}

Although we have unambiguously shown in Section~\ref{sec:Cu2+} that the minor signal of the (BEDT-TTF)$_2$ defect spins is distinct from those of the \ce{Cu^2+} impurities, some interaction is possible.
As illustrated in Fig.~1C of the main paper, the magnetic moments are defect spins in the BEDT-TTF layers due to defects among the valence bonds. The \ce{Cu^2+} impurities, on the other hand, are located within the anion sheets separating the organic layers perpendicular to the $bc$-plane.
From Fig.~\ref{fig:structure2} it is seen that these polymeric anions can be pictured as a hexagonal structure with copper ions in each corner. Due to intrinsic disorder in some of the linking cyano groups ($\cdots\ce{CN}\cdots$ {\it versus} $\cdots\ce{NC}\cdots$), the crystal symmetry is commonly assumed P2$_1$ \cite{Geiser91,Pinteric14,Dressel16}. The unit cell contains two formula units ($Z=2$) with two distinct configurations of copper, indicated in Fig.~\ref{fig:structure2}B.

In Fig.~\ref{fig:lowt_anisotropy}B we have seen that the angular dependence of the main signal is strongest for the $a^*c$-plane. This is explained by the spin-orbit interaction. Due to the tilt of the BEDT-TTF molecules with respect to the $bc$-plane, the maximum occurs approximately along the $a$-axis which is \SI{38}{\degree} off the $a^*$-direction. The defect signal has its extrema at the same angles in $a^*c$-plane as the main signal.  This is very much different along the other planes of rotation where the extrema of the main signal coincide with the crystallographic axes, but the defect signal exhibits an offset of \SI{22}{\degree} and \SI{25}{\degree}, respectively [Fig.~\ref{fig:lowt_anisotropy}B].

Let us suppose the defect spins are located exactly in the center of the (BEDT-TTF)$_2$ dimers. The angle dependence from Fig.~\ref{fig:lowt_anisotropy} imposes tight constraints on the orientation of dipole-dipole interactions: in particular, the axis connecting the defect spin on the (BEDT-TTF)$_2$ dimer with the searched-for local moment points at a considerable angle out of the layers. Hence, the $S=1/2$ entity that shifts the defect lines off the main signal is not located within the BEDT-TTF layers, i.e. the observed splittings are incompatible with dipolar coupling among neighboring organic dimers.
On the other hand, the $S=\frac{1}{2}$ defect spins might experience dipolar interaction with the \ce{Cu^2+} impurities in the anion layer, that also carry $S=\frac{1}{2}$. With two dimers per unit cell, there are four distinct distances to account for. These result in an additional local magnetic field of 1.6, 2.2, 3.5, and \SI{4.4}{mT}, respectively, that adds to the external field (the values listed here correspond to an alignment parallel to the axis connecting the two spins, i.e. $\alpha=0$ in Eq.~\ref{Eq-dipolar-coupling}).
In view of the experimentally observed local fields in the range of several mT (see Fig.~\ref{fig:lowt_anisotropy}), any interaction beyond the unit cell can be neglected due to the rapid $r^{-3}$ reduction with inter-spin distance.

The values associated with the nearest Cu positions are close to the shift of $\simeq 6$~mT away from the main line at an angle $22^{\circ}$ to the $c$-axis upon rotation within the $bc$ plane, see Fig.~\ref{fig:lowt_anisotropy}(A,B). We note that any structural modifications at $T^*$ may alter the distances and orientations between particular Cu sites and the center of (BEDT-TTF)$_2$ dimers, as compared to the structural data that were acquired at $T\simeq T^*$~\cite{Jeschke12}, but not significantly below the transition. This way, a change in Cu -- (BEDT-TTF)$_2$ distance of order 1~\AA\ or less can easily account for the observed local field of 6~mT. The fact that we observe several defect lines (at least 3) with slightly different angle dependence and field amplitude suggests that there are several possible Cu sites relevant for dipolar interaction with the defect spins on the organic dimers. Depending on growth conditions and lattice properties, some sites may occur more likely than others.

Moreover, the occurrence of \ce{Cu^2+} within the polymeric Cu$_2$(CN)$_3$ sheet likely causes a local distortion due to the additional charge. Given that the additional electron (required for charge neutrality of the crystal) is doped to the closest (BEDT-TTF)$_2$ dimer (which therefore acquires $S=0$), the resulting unpaired (BEDT-TTF)$_2$ spin is automatically located in the vicinity of the \ce{Cu^2+} site. This way, \ce{Cu^2+} impurities are very likely a relevant source of localized defect spins that contribute to the spin susceptibility in the otherwise spin-gapped crystal.

ESR measurements can reliably distinguish between (BEDT-TTF)$_2$ defect spins and \ce{Cu^2+} moments, yielding a pronounced drop of $\chi_S$ at $T^*$ even without separating the (BEDT-TTF)$_2$ main and defect signals~\cite{Komatsu96}. Bulk susceptibility measurements, on the other hand, measure the sum of all unpaired spins in the sample. Hence, adding also \ce{Cu^2+} to the (BEDT-TTF)$_2$ defect signal further conceals the opening of the spin gap in the intrinsic response. This can explain in part the weak and smeared-out drop in the SQUID measurements in Ref.~\cite{Shimizu03}. 

While \ce{Cu^2+} impurities seem a promising candidate, consistent with the experimental results and the crystal structure, we do not exclude other types of defects in the system. In Fig.~3(B), for instance, the low-temperature signal consists of the defect line discussed here and another absorption at the original position of the main line. In view of $T_{base}=0.025$~K~$\ll \Delta=12$~K, one can safely assume that at this temperature no valence bonds are thermally excited (i.e. spin singlets broken into two $S=1/2$). In addition, the observed signal with zero local field implies that there are no other unpaired moments in its near vicinity, suggesting that this type of defect spins emerges from a different source unrelated to \ce{Cu^2+}. Likely candidates are various types of intrinsic valence-bond defects such as domain boundaries, including the arrangement sketched in Fig.1(C) but not limited to that~\cite{Kimchi18,Liu18,Kawamura19}.

\clearpage

\section{Multi-frequency ESR spectroscopy using coplanar waveguide resonators}
\subsection{Experimental Setup and Procedure}
\label{sec:lowTsetup}
\begin{figure}[htb]
  \centering
  \includegraphics[width=0.6\linewidth]{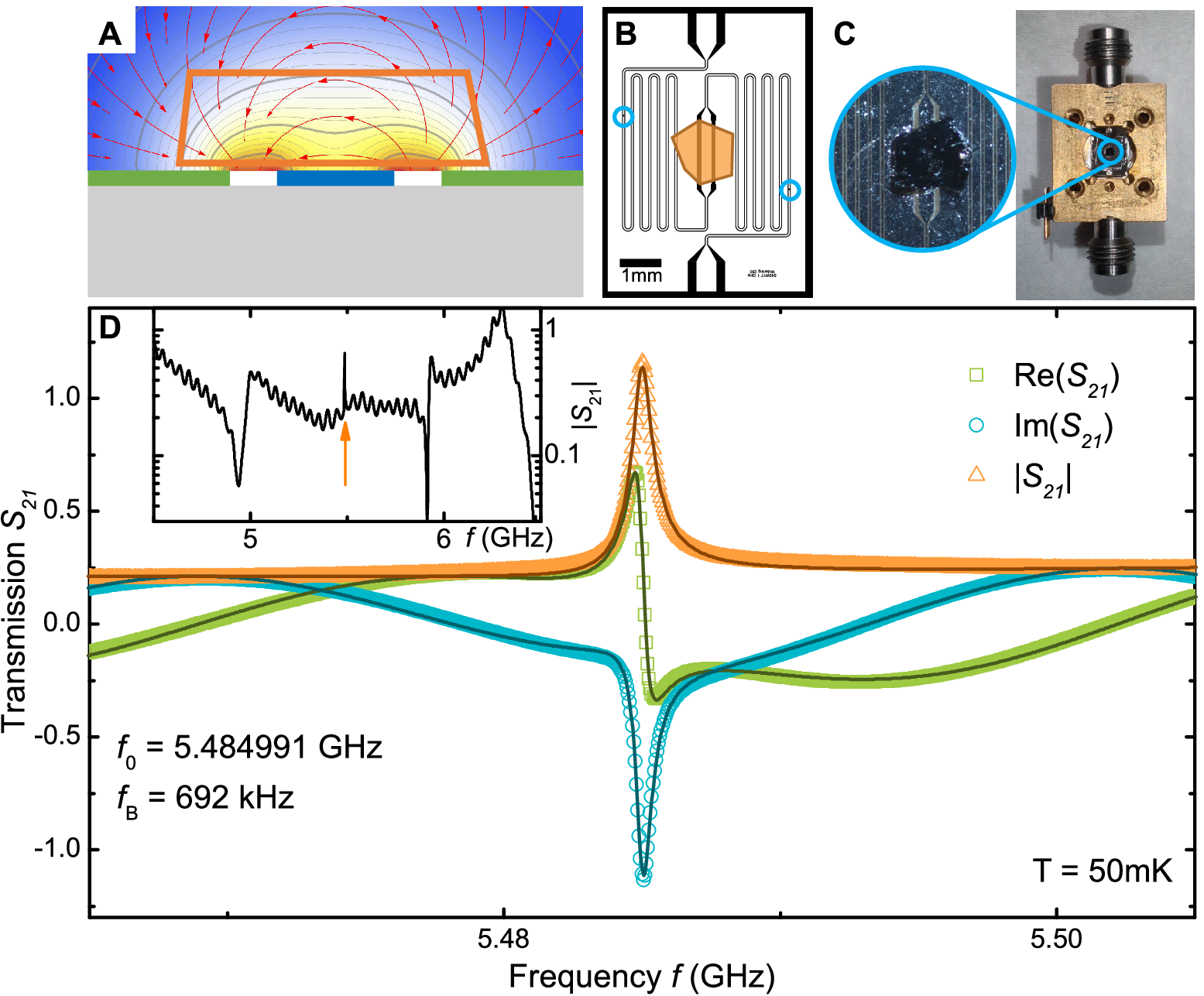}
  \caption{
  Coplanar waveguide resonators.
  (\textbf{A})~Cross-section of a coplanar waveguide with inner conductor (blue) and ground planes (green) on top of a substrate (grey). The microwave magnetic field (red field lines) encircles the inner conductor and interacts with a sample (orange box) placed on top of the metallization layer.
  (\textbf{B})~Layout of a coplanar wave\-guide resonator with a fundamental mode of \SI{1.1}{GHz}. The capacitive coupling gaps are indicated by blue circles.
  (\textbf{C})~Niobium coplanar waveguide mounted in its brass sample box with coaxial connectors. A sample is placed in the center of the resonator, as seen in the magnified spot.
  (\textbf{D})~Complex transmission spectra $S_{21}$ through the coplanar waveguide resonator in the vicinity of the $n=5$ mode at $B=0$. The lines are a fit by equation (\ref{eq:S21}). The center frequency is $f_0=\SI{5.48}{GHz}$, the width $f_B=\SI{0.69}{MHz}$, i.e.\ the quality factor of the loaded resonator $Q\approx 8000$. Inset: Absolute value of the transmission $|S_{21}|$ in a larger frequency range. Notice the sharp resonance compared to much broader background features.
  }
  \label{fig:planar_resonator}
\end{figure}

We employ coplanar waveguide resonators to measure electron spin resonance (ESR) at multiple frequencies down to mK temperatures. The details of the design, manufacturing, performance have been published in a series of articles over the past years \cite{Scheffler13,Clauss15,Voesch15,Javaheri16,Bondorf18,Rausch18,Miksch20}.
The chip with the metallic waveguide is mounted in a brass sample box (Fig.~\ref{fig:planar_resonator}C), which itself can be placed in a helium cryostat or a dilution refrigerator including a superconducting magnet.
This allows us to control the temperature $T>\SI{20}{mK}$ and apply a static magnetic field $B<\SI{8}{T}$.
The resonators are fabricated by optical lithography from a conducting thin film --~either metallic (copper or gold) or superconducting (niobium or YBCO)~-- on sapphire substrates.
A coplanar waveguide can be considered as a lengthwise slice of a coaxial cable. The center conductor as well as the ground plates separated by air gaps form a single plane.
Fig.~\ref{fig:planar_resonator}A shows the microwave magnetic field ${B}_1$ encircling the center conductor of a coplanar waveguide.
The magnetic field penetrates into the sample, which is placed on top of the waveguide, and can thus interact with the spins in the sample.
To enlarge the interacting volume, the waveguide can be structured in meander shape or increases in width as sketched in Fig.~\ref{fig:planar_resonator}B.
The standard ESR geometry is achieved with the static magnetic field ${B}$ parallel to the coplanar waveguide resonator  as the microwave magnetic field ${B}_1$ is then always perpendicular to ${B}$.
A one-dimensional resonator is formed between two capacitive coupling gaps in the coplanar waveguide resonator with modes at $n\lambda/2$ with $n\in\mathbb{N}$.

We measure the resonator modes with a vector network analyzer in transmission geometry.
Fig.~\ref{fig:planar_resonator}D shows a transmission spectrum of the real and imaginary part as well as the absolute value of $S_{21}(f)$ around the $n=5$ mode of a superconducting resonator with a fundamental mode at \SI{1.1}{GHz}.
Fitting the mode with a Lorentz function allows us to obtain the resonance frequency $f_0$ and bandwidth $f_B$.
Standing waves in the coaxial cables and spurious modes in the sample box lead to a background adding to the transmission of the ideal resonator in a complex way.
Therefore the fit of the resonator mode is conducted in the complex plane \cite{Thiemann18}.
Due to the high quality factor $Q\approx 3000 - 10\,000$, the resonances are typically very sharp compared to the background standing wave pattern as illustrated in the inset of Fig.~\ref{fig:planar_resonator}D, for instance.
This enables us to model the background up to linear order in the vicinity of the resonance frequency $f_0$.
We fit the resonance with the following model from \cite{Pozar12}:
\begin{equation}
  \label{eq:S21}
  S_{21}(f) = \exp\{\text{i}f\tau\} \bigg( \underbrace{ \frac{\nu_1}{f-\nu_2} }_{\text{resonance}}
 + \underbrace{ \nu_3+\nu_4\left(f-\Re(\nu_2)\right) }_{\text{background}} \bigg)
\end{equation}
Here the exponential prefactor accounts for the signal propagation time $\tau$ through the entire setup and $\nu_i$ are complex fit parameters, with the resonance frequency $f_0=\Re(\nu_2)$ and the bandwidth $f_B=2\Im(\nu_2)$.

\begin{figure*}[htb]
  \centering
  \includegraphics[width=0.7\linewidth]{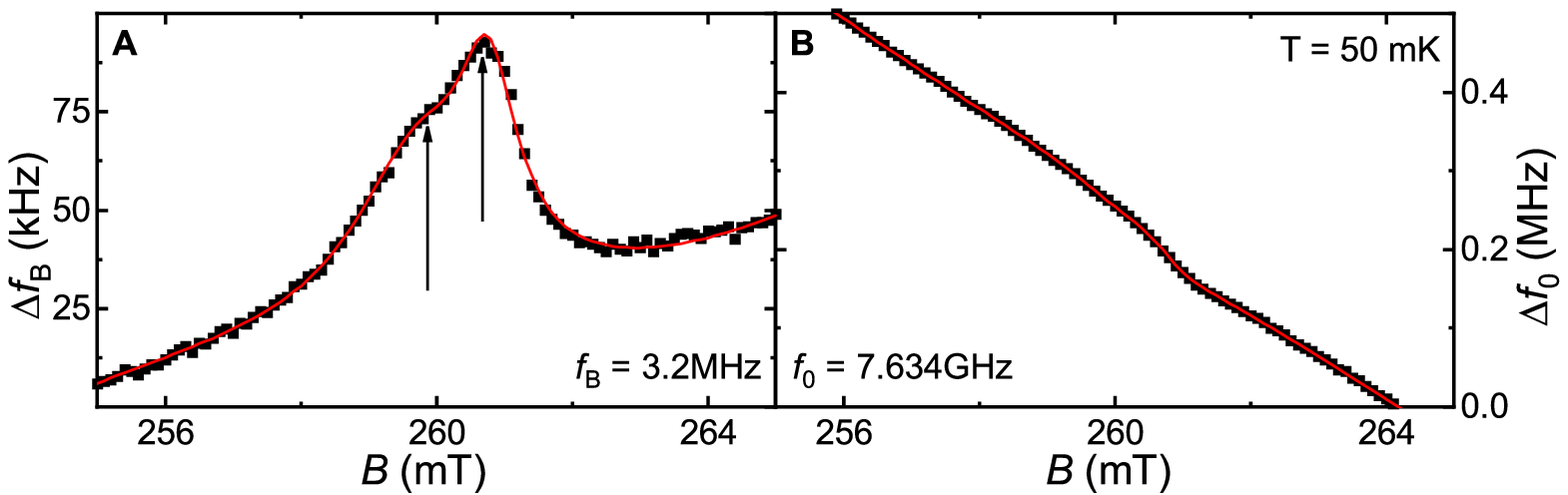}
  \caption{
  Low-temperature ESR spectra of \CuCN\ (sample \#\,1) from transmission measurements on a coplanar Nb resonator.
  (\textbf{A})~Fit of the bandwidth $f_\text{B}$ and (\textbf{B}) center frequency $f_0$ of the $n=7$ resonator mode
  affected by magnetic absorption. The arrows indicate the fields where the ESR absorption features are observed, corresponding to the strong main signal and the defect signal at lower fields. The field-dependent background, increasing $f_B$ and decreasing $f_0$, are due to suppression of superconductivity of the niobium resonator with increasing field.
  }
  \label{fig:planar_ESR}
\end{figure*}

To determine the ESR absorption, the external magnetic field $B$ is swept, and the transmission spectra are recorded at each field step, thus giving access to $f_0$ and $f_\text{B}$ as a function of $B$.
The sample with its complex susceptibility acts as a perturbation of the resonator.
The ESR dispersion $\chi^{\prime}$ and absorption $\chi^{\prime\prime}$ thus manifest as a change in the resonator frequency $\Delta f_0$ respectively the resonator bandwidth $\Delta f_B$ [see Fig.~\ref{fig:planar_ESR}].
Treating $\Delta f_0$ and $\Delta f_B$ as real and imaginary part of a complex quantity $\tilde{F}(B)$, we fit the obtained ESR spectra using $n=1$ or 2 Lorentzian functions for the $n=1$ or 2 resonances visible and a cubic polynomial background to fit the field dependent suppression of the superconductivity of the resonator
\begin{equation}
  \tilde{F}_\text{fit}(B)=\sum\limits_{k=0}^{3}\tilde{c}_k B^k + \sum\limits_{n}\frac{-\tilde{A}_n \Delta B_n}{(B - B_{\text{res},n}) + \text{i}\Delta B_n}
\end{equation}
Here the coefficients of the background polynomial $\tilde{c}_k$ as well as the amplitudes of the Lorentzians $\tilde{A}_n$ are complex quantities and thus fitted individually for the resonator frequency $\Delta f_0$ and the resonator bandwidth $\Delta f_B$. The resonance fields $B_{\text{res},n}$ as well as the linewidths $\Delta B_n$ are real numbers and therefore the same for dispersion respectively absorption.

The spectra shown in Fig.~\ref{fig:lowt_spectra_huebner}A as well as Fig.~3B show the quality factor $Q = f_0/f_B$ normalized by the fitted polynomial background.

\clearpage

\subsection{Frequency-dependent behavior of the defect signal for sample \#\,3}
\label{sec:lowT4}
\begin{figure}[ht]
  \centering
  \includegraphics[width=0.7\linewidth]{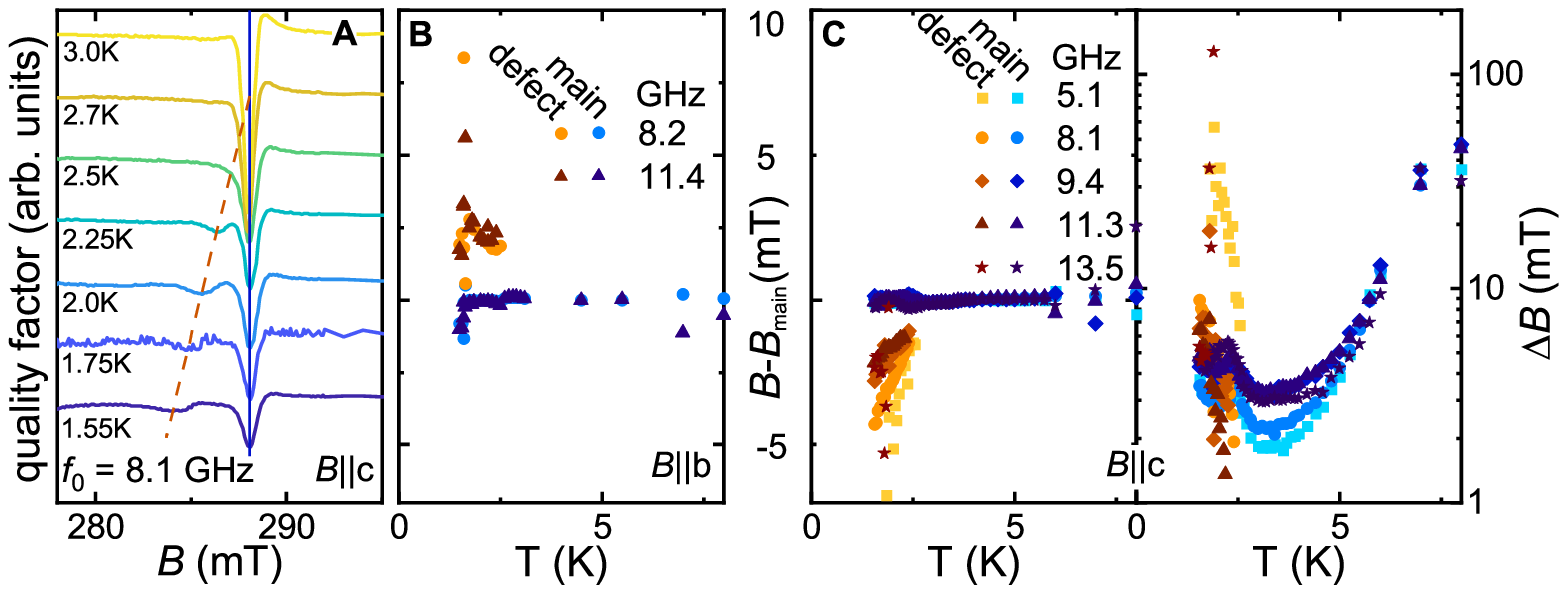}
  \caption{
  Emergence of the defect signal in \CuCN\ sample \#\,3 upon cooling.
  (\textbf{A})~Below $T_\text{loc}=\SI{2.5}{K}$ the defect signal (dotted line) becomes visible and shifts away from the main signal (dashed line) when cooled further.
  The resonance field of the signal at $B_\text{main}$ stays independent of frequency and temperature. The data are recorded for $B\parallel c$.
  (\textbf{B})~Temperature dependence of the resonance field when $B\parallel b$. Here the defect signal (orange) shifts to higher field compared to the main signal (blue).
  (\textbf{C})~Resonance shift of the defect signal (orange) for different frequencies; note $B\parallel c$. 
  The line width shows a minimum at around $T=\SI{3}{K}$.
  For lower temperatures the signals broaden, with the line width $\Delta B$ of the defect line being significantly larger than the line width of the main signal.
  }
  \label{fig:lowt_spectra_huebner}
\end{figure}

In order to illustrate the robustness of our findings and conclusions, and to give an impression about the sample-to-sample variation, in Fig.~\ref{fig:lowt_spectra_huebner} we present the results of frequency-dependent low-temperature ESR measurements on sample \#\,3, in the same way the findings on sample \#\,1 are shown in the main text, cf.\  Fig.~3.

Here we show the frequency and temperature dependent evolution of the defect signal for sample \#\,3 down to $T=\SI{1.5}{K}$.
The data has been recorded in a ${^4}$He cryostat using metallic coplanar waveguide resonators \cite{Javaheri18},
as described in Section~\ref{sec:lowTsetup}.
The behavior qualitatively matches the one shown for sample \#\,1 in Fig.~3 of the main paper.
The $g$-value of the defect signal is smaller compared to the main signal in case of $B||b$ and larger for $B||c$, in accord with the angular dependence shown in Fig.~\ref{fig:lowt_anisotropy}.
The frequency-dependent measurements presented in Fig.~\ref{fig:lowt_spectra_huebner}C exhibit a larger $g$-shift for smaller frequencies. The saturation of the $g$-value as well as the line width observed for sample \#\,1 below $T=\SI{1}{K}$ cannot be observed here as the temperature only reaches as low as \SI{1.5}{K}

\clearpage

\bibliographystyle{Science}
\bibliography{kCuCN_EPR}